\documentclass[pdflatex,sn-mathphys-num]{sn-jnl}

\usepackage{graphicx}
\usepackage{multirow}
\usepackage{amsmath,amssymb,amsfonts}
\usepackage{amsthm}
\usepackage{mathrsfs}
\usepackage[title]{appendix}
\usepackage{xcolor}
\usepackage{textcomp}
\usepackage{manyfoot}
\usepackage{booktabs}
\usepackage{algorithm}
\usepackage{algorithmicx}
\usepackage{algpseudocode}
\usepackage{listings}
\usepackage{bm}
\usepackage{float} 
\usepackage{subfigure}
\usepackage{makecell}
\usepackage{natbib}
\usepackage{color}
\usepackage{ulem}
\usepackage{soul}
\usepackage{wasysym}
\usepackage[utf8]{inputenc}

\pdfoutput=1

\theoremstyle{thmstyleone}

\theoremstyle{thmstyletwo}

\theoremstyle{thmstylethree}

\newcommand{\revise}[1]{\textcolor{black}{{#1}}}

\begin{document}

\title[Article Title]{Integrating Deep-Learning-Based Magnetic Model and Non-Collinear Spin-Constrained Method: Methodology, Implementation and Application}

\author[1]{\fnm{Daye} \sur{Zheng}}

\author[1]{\fnm{Xingliang} \sur{Peng}}

\author[1]{\fnm{Yike} \sur{Huang}}

\author[1]{\fnm{Yinan} \sur{Wang}}

\author[1]{\fnm{Duo} \sur{Zhang}}

\author[2,3]{\fnm{Zhengtao} \sur{Huang}}

\author[8]{\fnm{Zefeng} \sur{Cai}}

\author[4,1]{\fnm{Linfeng} \sur{Zhang}}

\author*[1,5]{\fnm{Mohan} \sur{Chen}}\email{mohanchen@pku.edu.cn}

\author*[6]{\fnm{Ben} \sur{Xu}}\email{bxu@gscaep.ac.cn}

\author*[1,7]{\fnm{Weiqing} \sur{Zhou}}\email{weiqingzhou@whu.edu.cn}

\affil[1]{AI for Science Institute, Beijing 100080, P. R. China}
\affil[2]{Graduate School of China Academy of Engineering Physics, Beijing 100088, P. R. China}
\affil[3]{International School of Materials Science and Engineering, Wuhan University of Technology, Wuhan 430070, P. R. China}
\affil[4]{DP Technology, Beijing 100080, P. R. China}
\affil[5]{HEDPS, CAPT, School of Physics and College of Engineering, Peking University, Beijing, 100871, P. R. China}
\affil[6]{Graduate School of China Academy of Engineering Physics, Beijing 100193, P. R. China}
\affil[7]{Key Laboratory of Artificial Micro- and Nano-structures of Ministry of Education and School of Physics and Technology, Wuhan University, Wuhan 430072, P. R. China}
\affil[8]{Department of Materials Science and Engineering, Carnegie Mellon University, Pittsburgh, PA 15213, USA}

\abstract{We propose a non-collinear spin-constrained method that generates training data for deep-learning-based magnetic model, which provides a powerful tool for studying complex magnetic phenomena that requires large-scale simulations at the atomic level. First, we propose a basis-independent projection method for calculating atomic magnetic moments by applying a radial truncation to numerical atomic orbitals. A double-loop Lagrange multiplier method is utilized to ensure the satisfaction of constraint conditions while achieving accurate magnetic torque. The method is implemented in ABACUS with both plane wave basis and numerical atomic orbital basis. We benchmark the iron (Fe) systems and analyze differences from calculations with the plane wave basis and numerical atomic orbitals basis in describing magnetic energy barriers. Based on an automated workflow composed of first-principles calculations, magnetic model, active learning, and dynamics simulation, more than 30,000 first-principles data with the information of magnetic torque are generated to train a deep-learning-based magnetic model DeePSPIN for the Fe system. By utilizing the model in large-scale molecular dynamics simulations, we successfully predict Curie temperatures of $\alpha$-Fe close to experimental values.}

\maketitle

\section{Introduction}{\label{sec:intro}}
The study of magnetic materials has been a cornerstone of condensed matter physics, with implications ranging from fundamental science to technological applications. Density Functional Theory (DFT)~\cite{1964-hk,1965-ks} has emerged as a powerful tool for understanding the electronic and magnetic properties of these materials~\cite{2023-misha}. However, accurate description of excited magnetic states within DFT has remained a formidable challenge due to the complex interplay between electron and lattice behaviors. 

The development of constrained Density Functional Theory (cDFT) has marked a significant advancement in this field ~\cite{1984-dederichs,2012-constrained-review}. In the pioneering work of Dederichs {\it et al.} in 1984~\cite{1984-dederichs}, DFT was extended to arbitrary constraints through the introduction of the Lagrange multiplier method. Since then, several cDFT methods have been used to study excited states of charge distribution~\cite{1984-dederichs,2005-wu} or magnetization~\cite{1998-stocks,2004-kurz,2018-SIESTA,2000-gebauer,2015-ma,2020-SPHInX,2023-cai}. Introducing penalty functions is commonly adopted to implement any constraint.~\cite{2000-gebauer,2015-ma}. Wu and Van Voorhis (2005)~\cite{2005-wu} introduced an efficient algorithm, which allows for the use of a double-loop method to find the effective Lagrange multiplier that satisfies the constraints. In addition, the cDFT methods have been implemented based on different basis sets, such as the real-space grid~\cite{2000-gebauer}, the full-potential linearized augmented plane-wave (FLAPW)~\cite{2004-kurz}, the projector augmented-wave (PAW)~\cite{2015-ma,2020-SPHInX,2023-cai}, and the numerical atomic orbitals (NAOs)~\cite{2018-SIESTA}. These efforts provide powerful tools in understanding charge and magnetization fluctuations in solids, predicting spin-dependent phenomena, and characterizing electron transfer reactions in molecules~\cite{2012-constrained-review,1998-stocks,2004-kurz,2005-wu,2015-ma,2018-SIESTA,2020-SPHInX,2023-cai}.

In recent years, the integration of deep learning models with DFT has offered unprecedented opportunities for transferring first-principles accuracy to larger scales~\cite{2024-dpa1,2023-dpa2,2024-Qi,2022-Chen,2023-Deng,2023-Batatia,2023-Choudhary,2021-choudhary}. The available databases~\cite{2013-mp,2020-materials-cloud,atomly,2024-OMat24} contain a large amount of DFT data that can be used to train machine learning models for different purposes.~\cite{2024-unisemi,wu2023universal,Wu23p144102}. Recently, some magnetic models~\cite{2024-fe-npj,2024-deepspin} have emerged that incorporate the degrees of freedom of spin into the existing neural networks and can perform large-scale magnetic dynamics simulations. However, these models suffer from a lack of data since most existing DFT data in datasets are non-magnetic or only include ground-state magnetic results. Ground-state data cannot meet the sampling requirements of magnetic models with an additional degree of freedom spin. From this perspective, spin-cDFT methods not only act as powerful first-principles tools for studying magnetic excited states at the atomic scale but also provide training data for machine-learning-based magnetic models. 

To address the substantial data requirements for AI-driven magnetic modeling, cDFT methodology must simultaneously satisfy multiple critical requirements: (1) robust convergence and high accuracy for reliable large-scale sampling, (2) consistent precision across datasets to eliminate untraceable errors, (3) minimal dependence on manually tuned parameters for automated active learning workflows, (4) adaptable constraint conditions accommodating diverse magnetic systems, and (5) optimized computational efficiency to address the inherent cost premium of magnetic versus non-magnetic calculations. To establish a robust data generation framework meeting these criteria, we have developed and implemented a Lagrange multiplier-based cDFT approach within the open-source ABACUS software~\cite{Chen2010,Li2016,lin_accuracy_2020,2025-zhou}, which provides energies, atomic forces, stresses, and magnetic torques for any magnetic excited state, making it a suitable data engine for magnetic models. Additionally, we propose a basis-independent local orbital projection method for calculating magnetic moment magnitudes. Unlike the Mulliken population method~\cite{2018-SIESTA}, which, while maintaining the sum rule, may introduce contributions from non-local orbitals. Benefiting from the locality of NAOs, magnetic moments are obtained through the projection of atomic orbitals. By appropriately truncating the local atomic orbitals and smoothing near the truncation, we find that the modulated atomic orbitals provide satisfactory performance in stability and convergence during cDFT calculations. We present a unified implementation of cDFT that works efficiently with both plane-wave (PW) and numerical atomic orbital (NAO) basis. The spin-cDFT is fully optimized using the double-loop approach~\cite{2005-wu}. The optimized torque holds significant value for magnetic dynamics simulations and configuration space exploration. While conventional penalty methods~\cite{2015-ma} yield approximate torques $\lambda$, their fixed $\lambda$ induces non-negligible errors in energy derivatives. The adaptive scheme circumvents this by enforcing $\Delta M \to 0$. Moreover, the penalty methods encounter convergence difficulties when $\lambda$ is too large~\cite{2020-streib-prb}, which requires careful handling of the iterative process. More critically, while the penalty function method can achieve reliable calculations for case studies through careful adjustment of $\lambda$, this system-dependent tuning process becomes impractical for the AI-driven large-scale sampling. A predetermined $\lambda$ exhibits varying accuracy levels across different configurations, introducing untraceable error that compromise subsequent model training. The limited accuracy, unstable convergence, and untraceable errors in large datasets render the penalty function method inadequate for meeting requirements 1/2/3. In contrast, the Lagrange multiplier method effectively addresses these challenges through its dual-optimization scheme~\cite{2004-kurz,2005-wu,2020-SPHInX,2023-cai}.

Our implementation supports two optional constraint modes: the first constrains both the magnitude and direction of magnetic moments, while the second constrains only the moment direction. Despite Gyorffy {\it et al.}'s assertion that longitudinal and transverse spin fluctuations exhibit temporal separation~\cite{1985-gyorffy-first}, recent studies indicate that longitudinal spin fluctuations play a significant role in processes such as ferromagnetic-paramagnetic transitions~\cite{2007-ruban-prb} and phonon-magnon interactions~\cite{2018-wu-magnon}. In this work, we primarily focus on the application of the full-constraint algorithm, enabling the construction of magnetic potential energy surfaces through integration with recently developed magnetic models~\cite{2024-deepspin}. In coupled spin-lattice dynamics~\cite{2025-huang-prep}, the contribution of longitudinal spin fluctuations manifests in the thermodynamic statistical properties of equilibrium states across different temperatures within the energy domain. The real-time evolution of longitudinal fluctuations in the temporal domain falls outside the scope of the present study.

Through the synergistic integration of cDFT calculations, magnetic model, active learning algorithms, and coupled spin-lattice dynamics~\cite{2025-huang-prep}, we have developed a comprehensive end-to-end framework for autonomous magnetic model development. The workflow achieves complete automation throughout the entire training process, eliminating the need for manual intervention. Starting from scratch, we successfully trained two distinct models using PW and NAO basis, respectively. Both models yielded quantitatively consistent ferromagnetic-paramagnetic phase transition. This robust agreement across independent models provides validation of our methodology's accuracy and transferability. This research paradigm offers valuable insights for AI-driven magnetic materials simulation. Furthermore, our work provides the research community with a reliable and accessible tool for related investigations.

The content of this paper is as follows. Sec.~\ref{sec:method} details the theoretical foundations, including non-collinear spin, projection methods, and spin-constrained DFT. Sec.~\ref{sec:proj_test} discusses the modulated NAOs, as well as the behavior of atomic magnetic moments based on orbital projections. Sec.~\ref{sec:implementation} verifies the correctness of the implementation through finite difference tests. Sec.~\ref{sec:Fe} introduces the properties of various magnetic excited states of pure Fe calculated using spin-cDFT. Sec.~\ref{sec:deepspin} employs iron (Fe) as a prototype system to demonstrate our automated workflow that bridges first-principles cDFT calculations with AI magnetic model, and provides an accurate Curie temperature through molecular dynamic simulation. Finally, Sec.~\ref{sec:conclusion} summarizes the work.

\section{Results}

\subsection{Projection Methods} \label{sec:proj_test}

\begin{figure*}
    \centering    
    \includegraphics[width=1.0\textwidth]{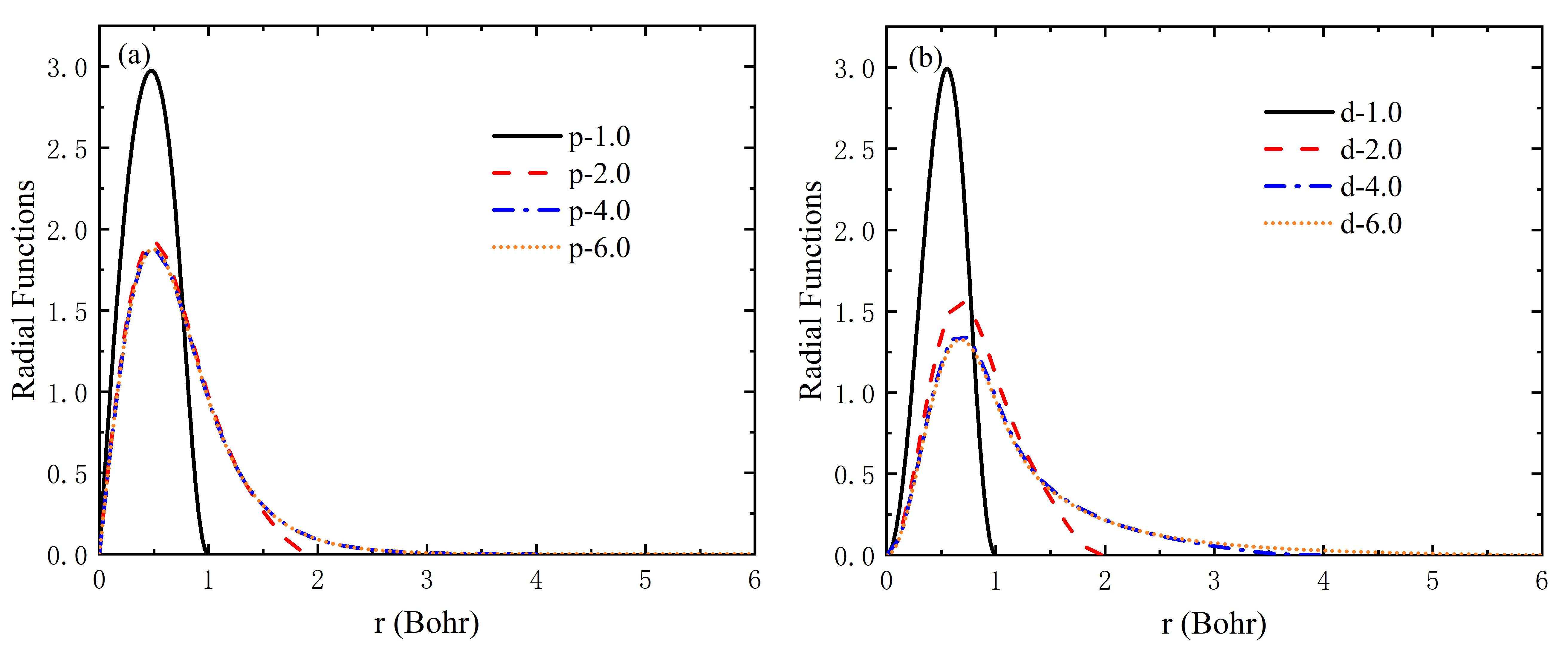}
    \caption{\textbf{Radial functions (RFs) of the first-$\zeta$ numerical atomic orbital of Fe before and after modulation.} (a) The original and modulated RFs for $p$ orbitals, where the original numerical atomic orbitals have a cutoff radius $r_c$ of 6.0 Bohr, as well as the modulated orbital with the modification radius $r_m$ as 1.0, 2.0 and 4.0 Bohr. ``p-1.0'' represents the $p$ orbital modulated by Eq.~\ref{eq:orb_mod} with $r_m=1.0$ Bohr. (b) The original and modulated RFs for $d$ orbitals.}
    \label{fig:orbital_modulate}
\end{figure*}

We have developed an innovative algorithm designed to estimate and control atomic magnetic moments using localized orbital projection techniques (see methods in Sec.~\ref{sec:proj_method}). The key innovation of our method lies in the localized modulation of the numerical atomic orbital basis. We have benchmarked this algorithm on ferromagnetic (FM) and antiferromagnetic (AFM) iron (Fe) bulk.

Our first step involved comparing the modulated radial functions (RFs) of orbitals obtained through our algorithm with the original NAO RFs. It is important to distinguish the two properties discussed here: the cutoff radius $r_c$ and the modulation radius $r_m$ of the orbitals. The original NAO is zero beyond the cutoff radius, while the modulation radius refers to the range of the original NAO modulated by Eq.~\ref{eq:orb_mod} for magnetic moment projection. In Fig.~\ref{fig:orbital_modulate}, we modify the shape of a NAO using Eq.~\ref{eq:orb_mod} with $r_m$ ranging from 1 to 5 Bohr, where the $r_c$ of original NAO is 6.0 Bohr. As shown in Fig.~\ref{fig:orbital_modulate}(a), the modulated $p$ orbitals align closely with the original orbital shape if $r_m \geq 2$ Bohr, while the modulated $d$ orbitals remain consistent with the original shape for $r_m \ge 3$ Bohr. These modulated orbitals exhibit smoothness at $r_m$, with the first derivative approaching zero. For more localized $r_m$, the modulated orbitals retain the peak radius of the RFs, but their peak heights are significantly increased due to normalization constraints. Excessively large $r_m$ may introduce non-localized electron wavefunctions or portions of wavefunctions from localized orbitals of other atoms into the projection results, thereby preventing the projection from accurately representing atomic localized orbitals. On the other hand, excessively small $r_m$ can lead to significant distortion, as they cause a large deviation from the original atomic orbitals.

Next, we discuss the criteria for selecting appropriate $r_m$. To this end, we calculated the ground state of BCC phase iron (BCC-Fe) for both ferromagnetic (FM) and anti-ferromagnetic (AFM) configurations using PW and NAO basis sets. The atomic magnetic moments of Fe atom were then estimated using the modulated orbital local projection algorithm with various $r_m$. As a reference, we define the ``TMAG'' by directly summing the magnetic density within the cell $\sum_{\mathbf{r}}\mathbf{m(r)}/N$ for the FM configuration, and ``AMAG'' as the modulus sum of the magnetic density $\sum_{\mathbf{r}}|\mathbf{m(r)}|/N$ for the AFM configuration, where $N$ is the number of Fe atom. For the NAO calculations, we can use the high-precision TZDP-10 Bohr results as a AFM reference.

\begin{figure*}
    \centering    
    \includegraphics[width=1.0\textwidth]{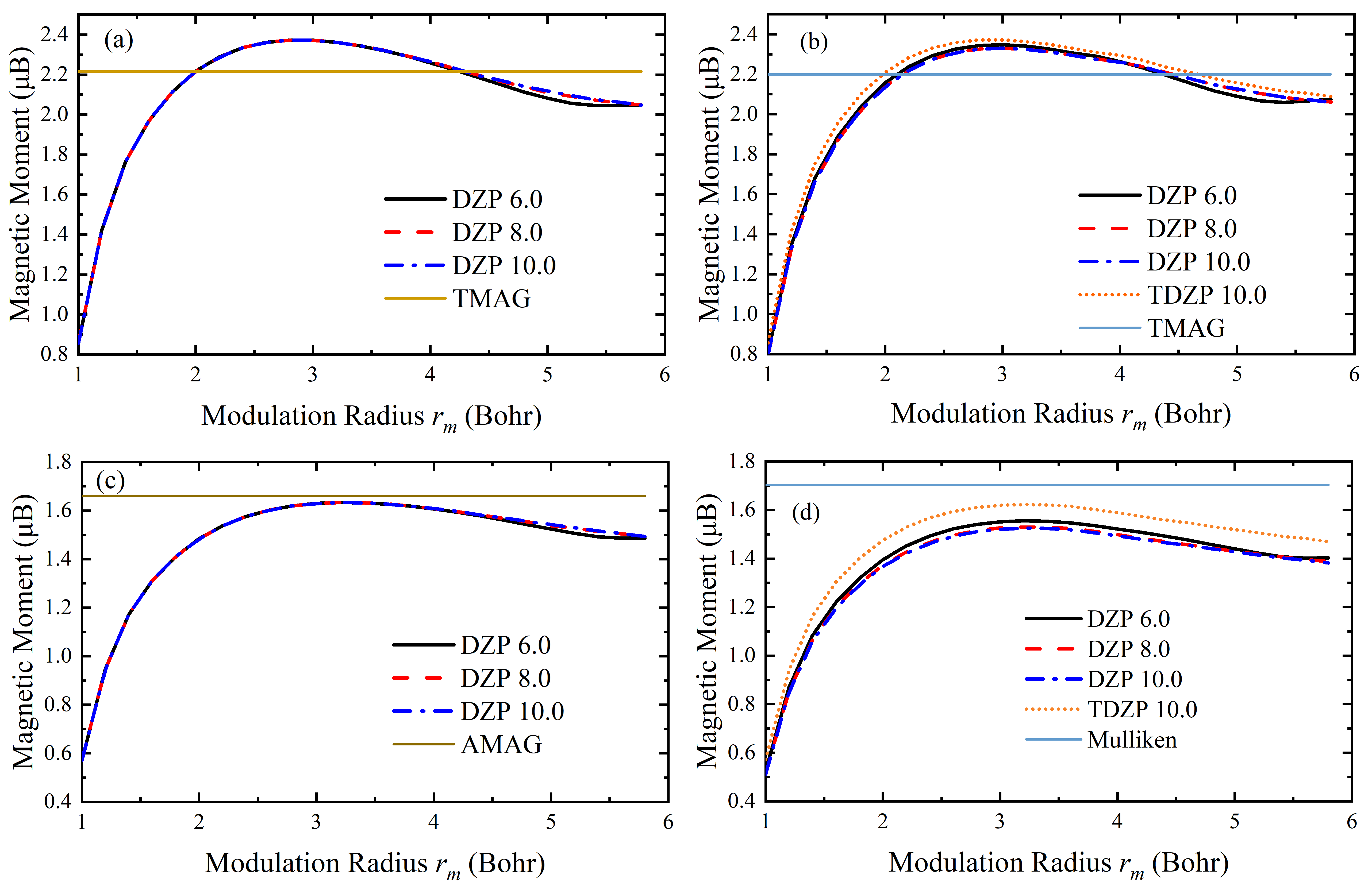}
    \caption{\textbf{Estimation of BCC-Fe atomic magnetic moments as a function of modulation radius $r_m$.} The modulated orbital projection algorithm is employed with DZP and TZDP basis sets with cutoff radius $r_c$ ranging from 6.0 to 10.0 Bohr, to demonstrate the impact of different NAO basis sets on atomic magnetic moments. Where (a) is for the PW basis set in the FM magnetic configuration, with the reference value being the total magnetization per atom (TMAG); (b) is for the NAO basis set in the FM magnetic configuration, with the reference value being the TMAG; (c) is for the PW basis set in the AFM magnetic configuration, with the reference value being the absolute magnetization per atom (AMAG); and (d) is for the LCAO basis set in the AFM magnetic configuration, with the reference value being the atomic magnetization from Mulliken charge.}
    \label{fig:projected_mag}
\end{figure*}

As shown in Fig.~\ref{fig:projected_mag}(a) and (c), the projected magnetic moments obtained from the different modulated orbitals agree well with each other if $r_m < 4$ Bohr in all PW calculations. This indicates that the projection results are dependent solely on the selected $r_m$, rather than on $r_c$. The results from the NAO basis (Fig.~\ref{fig:projected_mag}(b) and (d)) show differences in the projection results between various basis sets, with these discrepancies arising solely from the NAO basis set itself. Furthermore, the comparison between left and right panel of Figs.~\ref{fig:projected_mag} shows the higher-precision TZDP results align well with the PW ones. Fig.~S8 further compares the projected magnetic moments versus modulation radius $r_m$ across different basis sets (PW, SZ, DZP, TZDP). For BCC-Fe ferromagnetic state calculations, the maximum errors are 0.13 $\mu_B$ (SZ), 0.03 $\mu_B$ (DZP), and 0.002 $\mu_B$ (TZDP).

\begin{figure*}
    \centering    
    \includegraphics[width=1.0\textwidth]{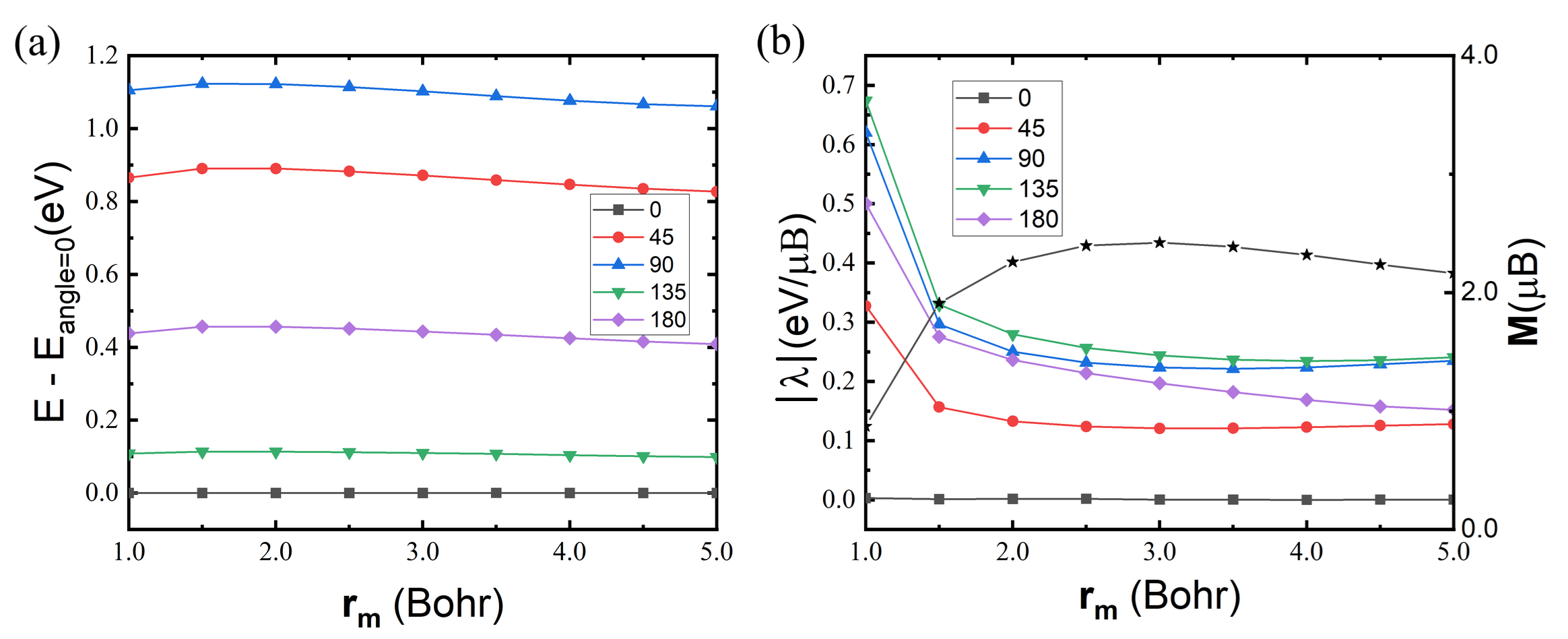}
    \caption{\textbf{The total energy and magnetic torque of BCC-Fe as function of modulation radius.} (a) The total energy of BCC-Fe calculated at different modulation radius by using DZP basis ($r_c=7.0~\text{Bohr}$). The labels represent the angle of magnetic moment between two nearest-neighbor Fe atoms in BCC-Fe, while the magnetic moment magnitudes are set to the corresponding FM ground-state values for each $r_m$, as indicated by the black pentagrams in (b). (b) shows the corresponding magnetic torque $|\boldsymbol{\lambda}|$ of BCC-Fe calculated at different modulation radius.}
    \label{fig:rcut-E-lambda}
\end{figure*}

The choice of an appropriate modulation radius $r_m$ significantly affects the estimated values of atomic magnetic moments. We discuss two possible strategies for determining $r_m$. A direct approach is to calibrate the projected orbital radius by matching it to reference values, such as experimental data or literature values. For BCC-Fe, the modulation radius chosen using the first approach is 2.0 Bohr, with the FM and AFM atomic magnetic moments being 2.21 and 1.48 $\mu_B$, respectively. These values are in good agreement with those calculated by VASP~\cite{2015-ma}, which are 2.22 and 1.52 $\mu_B$. However, this can be challenging for complex compounds or systems without available experimental data, as it may be difficult to identify suitable reference values.

One application of our implementation is to generate a large-scale dataset for subsequent model training. The dataset required for the model includes the system energy $E$, atomic magnetic moments $\mathbf{M}$, and the magnetic torque $\boldsymbol\lambda$. It should be clarified that the Lagrange multipliers in this work differ fundamentally from those in other implementations~\cite{2015-ma,2020-SPHInX}, as their physical interpretation is intrinsically determined by the specific constraint conditions (see Sec.~\ref{sec:spin-cDFT} for the definition of $\boldsymbol{\lambda}$ in this work). The influence of the modulation radius on all these physical quantities should be taken into consideration. Here we present the calculated energy and magnetic torque for BCC Fe at different modulation radius in Fig.~\ref{fig:rcut-E-lambda}. It can be observed that within a certain range (1-5 Bohr), different $r_m$ have a small quantitative impact on the energy, and do not lead to qualitative changes, such as alterations in the energy ordering of different states. Compared to the energy, the influence of the $r_m$ on the magnetic torque is more pronounced. 
\revise{When the modulated orbital cover only an extremely narrow region near the nucleus (1-2 Bohr), truncation distorts the wavefunction characteristics of the original NAOs. To achieve the same $\mathbf{M}_{\rm{target}}$ within a reduced integration volume, a stronger constraint field must be applied, causing $|\boldsymbol\lambda|$ to rise sharply. In contrast, when $r_m$ is sufficiently large to encompass the main valence characteristic peak of the NAOs, the projector exhibit high spatial overlap with the true orbitals. Here, $\boldsymbol\lambda$ only needs to compensate for minor deviations between the self-consistent electron distribution and the target magnetic moment, and its value stabilizes. Fig.~\ref{fig:rcut-E-lambda}b shows that $|\boldsymbol\lambda|$ remains essentially constant beyond $r_m \geq 3$ Bohr, indicating that the physical quantity has become independent of the projection parameter.}
We wish to emphasize that for an individual calculation, these physical quantities are self-consistent, as guaranteed by the finite difference tests will be presented below. However, considering high-throughput calculations and the further exploration of sample space at high temperatures, one can anticipate the appearance of a large number of structurally complex configurations. We are concerned that in highly non-uniform configurations, even minor disturbances may lead to significant changes in physical quantities, potentially causing difficulties in model training. Based on these considerations, we aimed to find $r_m$ where the physical quantities exhibit relatively smooth behavior with respect to changes in real-space coordinates, that is $\frac{\partial E}{\partial r_m} \approx 0,~\frac{\partial \mathbf{M}}{\partial r_m} \approx 0,~\frac{\partial \boldsymbol \lambda}{\partial r_m} \approx 0$. Based on Fig.~\ref{fig:projected_mag} and Fig.~\ref{fig:rcut-E-lambda}, we selected 3.0 Bohr as the setting for producing the dataset in this work. This $r_m$ estimated atomic magnetic moments being 2.36 $\mu_B$ for FM and 1.6 $\mu_B$ for AFM in BCC Fe. 
\revise{By comparing with the direction of the total magnetic moment, this projection scheme is also demonstrated to accurately maintain alignment between the target magnetic moment and the self-consistent spin directions (see Table.~S5).}
We wish to clarify that the different magnetic torques generated by different $r_m$ do not introduce systematic bias into the magnetic dynamics simulations for which the model is ultimately intended. This point is demonstrated in Fig.~S10 of the Supplementary Information.

\subsection{Finite Difference Tests for Spin-cDFT Method} \label{sec:implementation}

\begin{figure}
    \centering    
    \includegraphics[width=0.7\linewidth]{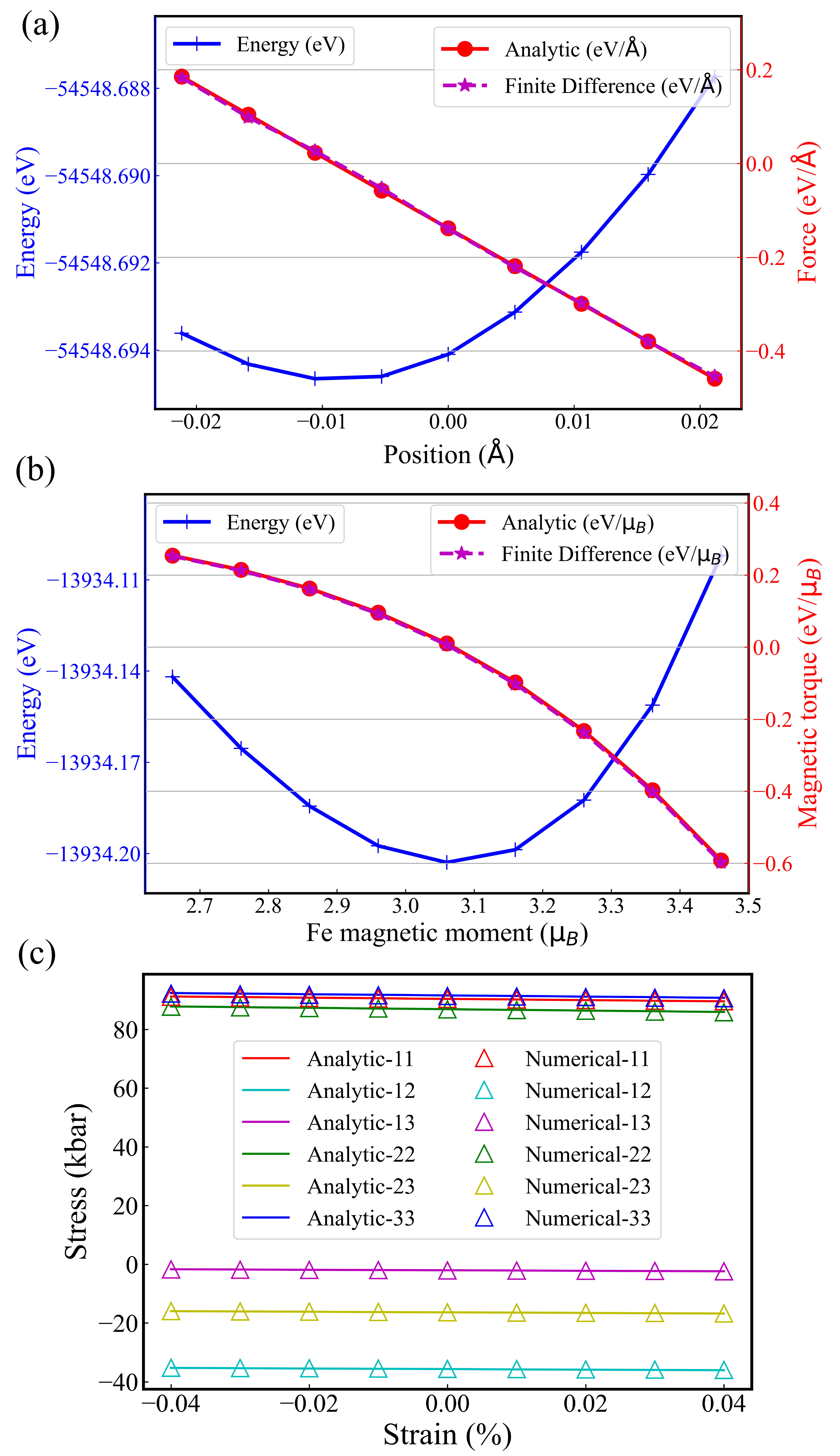}
    \caption{\textbf{Finite difference tests.} (a) Finite difference tests for the atomic force in BCC Fe$_{16}$ with a perturbation step of 0.01 Bohr along z-direction. (b) Finite difference tests for the magnetic torque in binary alloy FePt with a perturbation step of 0.1 $\mu_B$. (c) Finite difference tests for the cell stress in ternary alloy NiMnTi with a perturbation step 0.0001. ``11'' refers to the component of stress matrix $\sigma_{11}$. In (a-c), ``Analytic'' represents the value calculated from the formula, while ``Numerical'' is the value determined from the finite difference method.}
    \label{fig:FD}
\end{figure}

Based on the theory defined in Section~\ref{sec:method}, we implemented the spin-cDFT method in the open-source software ABACUS, the method is available with either plane wave or numerical atomic orbital basis sets. In the spin-cDFT framework, we introduced corrections for the energy, atomic forces, and lattice stresses under full magnetic constraints and incorporated calculations of magnetic torque $\bm{\lambda}$. These magnetic torques are optimized in the inner loop of the self-consistent field iterations. To validate our implementation, we compared analytical solutions for atomic forces, lattice stresses, and magnetic torques against numerical ones obtained through finite difference method. As shown in Fig.~\ref{fig:FD}, these tests were conducted on elemental body-centered cubic (BCC) iron (Fe), FePt binary alloy, and NiMnTi ternary alloy. For a 16-atom BCC-Fe supercell, we perturbed one Fe atom's position, where the maximum discrepancy between numerical forces and analytical forces across all perturbations does not exceed 6 meV/$\rm \AA$. Magnetic torques are the partial derivatives of the energy with respect to magnetic moments, rather than atomic positions. Selecting one Fe atom in the FePt alloy, we perturbed its magnetic moment. The maximum error in the finite-difference values does not exceed 0.006 $\text{eV}/\mu_B$, where the analytical result is obtained from the inner optimization. For the stress tests, we applied varying magnitudes of lattice strain to NiMnTi and obtained numerical stresses through finite-difference calculations. All components of these numerical stress tensors showed discrepancies within 0.25 kpar when compared with analytical values, demonstrating good consistency. For these three tests, the cutoff energy was set to 100 Ry, and the Brillouin zone was uniformly sampled by $5\times5\times5$ Monkhorst-Pack grid. The raw data for all finite-difference tests are provided in Table.~S1/S2/S3 of Supplementary Information. The successful finite difference tests corroborate the proper functionality of the cDFT features for energy, atomic forces, magnetic torques, and lattice stress in ABACUS. 

\subsection{Magnetic Constraints and Energy Surface of Iron Phases} \label{sec:Fe}

\begin{figure*}
    \centering    
    \includegraphics[width=1.0\textwidth]{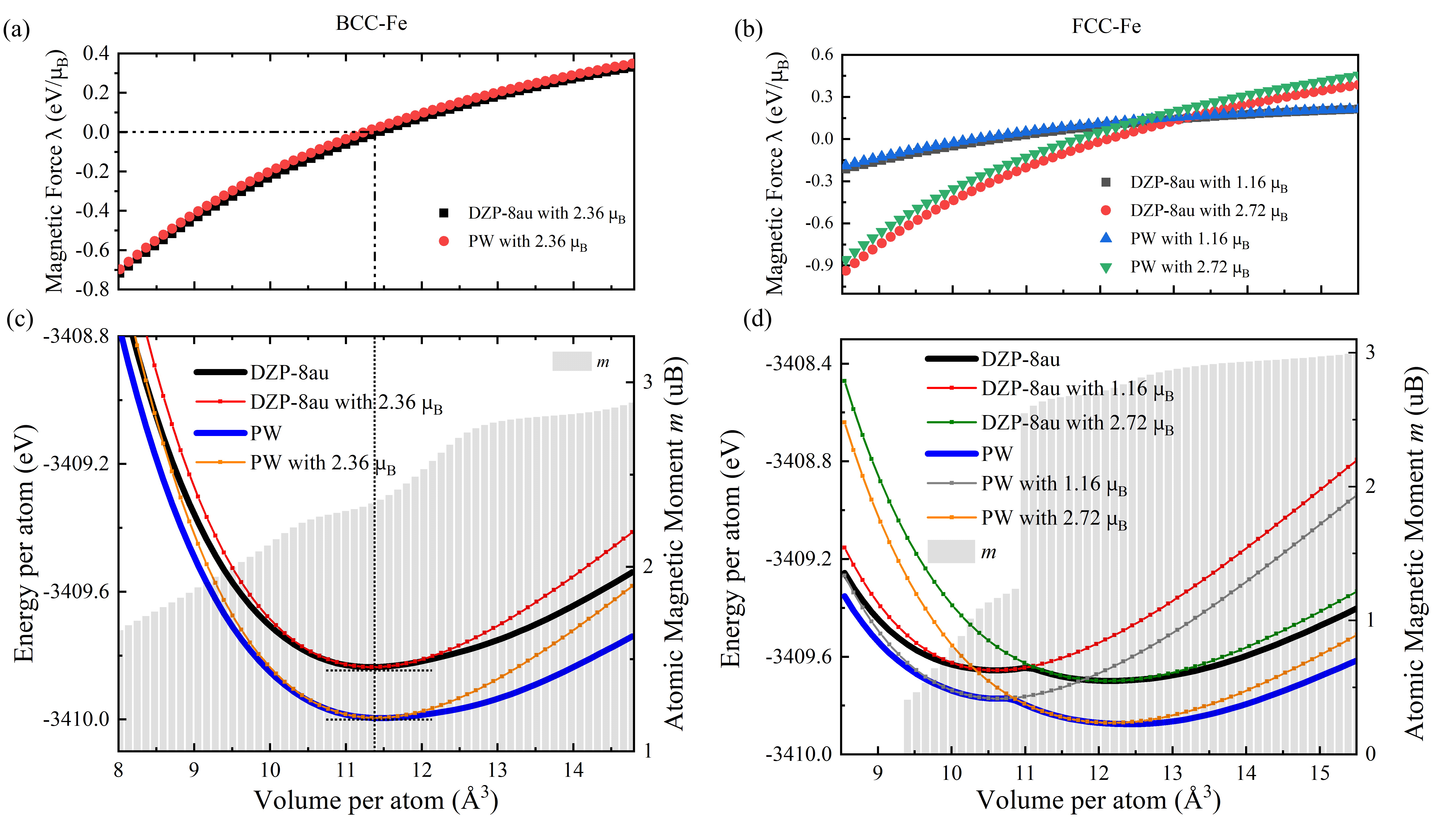}
    \caption{\textbf{Comparison between the unconstrained and constrained calculations.} (a,b) plot the magnetic force $\lambda=|\boldsymbol\lambda|$ as a function of the cell volume per atom for BCC-Fe (a) and FCC-Fe (b). (c,d) The solid lines show the total energy per atom as a function of the cell volume per atom for BCC-Fe (c) and FCC-Fe (d). Here ``DZP-8au-v2.1'' represents the v2.1 NAOs calculations based on the DZP orbitals with the 8 Bohr cutoff. The gray histograms represent the atomic magnetic moment $m$ after fully unconstrained self-consistent calculations. The dotted lines show the cDFT energy with certain constrains.}
    \label{fig:Fe_1D}
\end{figure*}

The spin-cDFT method presented in this paper facilitates calculations on any magnetic configurations and supports the analysis of complex magnetic structures. We take the bulk iron as an example to demonstrate the reliability of the magnetic constraint method in ABACUS. 

The BCC and FCC phases are two prevalent iron crystal structures. The BCC structure is the stable phase of iron at room temperature and is known for its high strength and low ductility. The BCC Fe structure exhibits a ferromagnetic (FM) state with atomic magnetism around 2.2 $\mu_B$ per atom~\cite{acet1994high}, and the Curie temperature was found at 1043 K experimentally~\cite{2001-bcc-fe}. On the other hand, The FCC structure has a higher density of atoms compared to BCC. Experimentally, iron undergoes a structural phase transition from BCC to FCC around 1085 K~\cite{2016-alloy-phase-diagram} and the FCC structure is the stable phase until 1667 K~\cite{2016-alloy-phase-diagram}. Although FCC Fe exhibits paramagnetic behavior in the experiment, theoretically, it is predicted that the energy of either antiferromagnetic (AFM) or double-layer antiferromagnetic (DAFM) states would be lower than that of ferromagnetic at 0 K~\cite{2024-fe-npj}. It has a more unimodal density of states at the Fermi level, which results in lower magnetic moments compared to BCC Fe.

\begin{figure*}
    \centering    
    \includegraphics[width=1.0\textwidth]{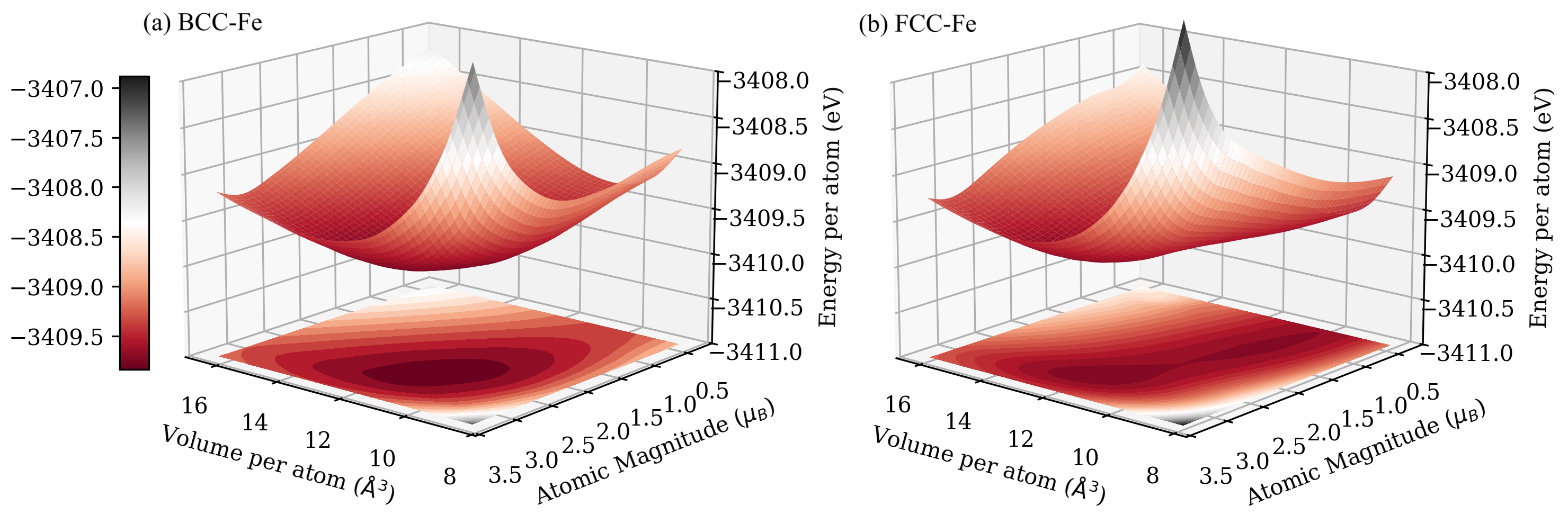}
    \caption{\textbf{Two-dimensional energy surface.} (a) The total energy of BCC-Fe as a function of atomic magnetic moment and cell volume. (b) The total energy of FCC-Fe as a function of atomic magnetic moment and cell volume. All calculations are performed by spin-cDFT with DZP-8au-v2.1 NAOs orbitals.}
    \label{fig:Fe_2D}
\end{figure*}

Firstly, we consider ferromagnetic BCC-Fe. In the lower panel of Fig.~\ref{fig:Fe_1D}(a), the blue solid line illustrates the total energy's dependence on the BCC-Fe cell volume, calculated using the PW basis without spin constraint. It is evident that the equilibrium volume of BCC-Fe, determined by the volume at which the total energy is minimized, is 11.2 \AA$^3$. The black line represents results computed with the v2.1 NAOs basis, specifically the DZP basis with a cutoff radius of 8 au, referred to as ``DZP-8au-v2.1''. The NAOs basis is optimized on the basis of the results from the PW basis set, and the energy difference between them reflects the quality of the NAOs orbitals. We observe that the energy difference remains nearly constant under tensile conditions but decreases significantly under compression, indicating discrepancies in orbital precision at varying interatomic distances. The grey histograms in the figure depict the atomic magnetic moments at different volumes. We present only the NAO results since the magnetic moments calculated using the PW basis set and the NAOs basis set are relatively similar. The atomic magnetic moments increase monotonically with volume, with the projected atomic magnetic moment reaching 2.36 $\mu_B$ at the energy minimum.

We now impose a constraint on the magnetic configuration, fixing the magnetic moments at 2.36 $\mu_B$. The corresponding energy from the spin-cDFT calculations is shown as dotted lines. Due to this constraint, the magnetic moment of 2.36 $\mu_B$ represents an excited state for all volume points except 11.2 \AA$^3$, resulting in energies higher than the unconstrained ground-state energy. The spin-cDFT energy intersects the ground-state energy at only one point with the magnetic moment 2.36 $\mu_B$. The upper panel of Fig.~\ref{fig:Fe_1D}(a) displays the magnetic torques $\bm{\lambda}$ optimized in the cDFT calculations. The magnetic torques are zero when the cDFT states are the same as the ground state, indicating that no penalty is required to maintain the specified magnetic moment configuration. However, the excited magnetic states lead to finite magnetic torque, which can be interpreted as an additional effective magnetic field necessary to constrain the magnetic moment.

Similarly, in Fig.~\ref{fig:Fe_1D}(b), we present the results for ferromagnetic FCC-Fe. Compared to BCC-Fe, the FCC phase near the ground state exhibits two local minima in the cell volume per atom, located around 10.5 \AA$^3$ and 12.2 \AA$^3$, respectively. The atomic magnetic moment increases monotonically with the cell volume. A key difference arises at the boundary between the two minima, where a sudden change in the magnitude of the atomic magnetic moment is observed. The atomic magnetic moments corresponding to the two minima are 1.16 $\mu_B$ and 2.72 $\mu_B$, respectively. We constrained the magnitude of the atomic magnetic moments to these two values, maintaining their direction in the ferromagnetic state. The results show that, regardless of whether the calculations are performed using the PW or NAO basis, the total energy under the constrained conditions only touches the ground-state energy at the corresponding ground-state magnetic moments, with the associated magnetic force $\boldsymbol\lambda$ being zero. When the constrained magnetic moments deviate from the ground-state values, the total energy exceeds the ground-state energy, and the magnetic force progressively increases.

\begin{figure*}
    \centering    
    \includegraphics[width=1.1\textwidth]{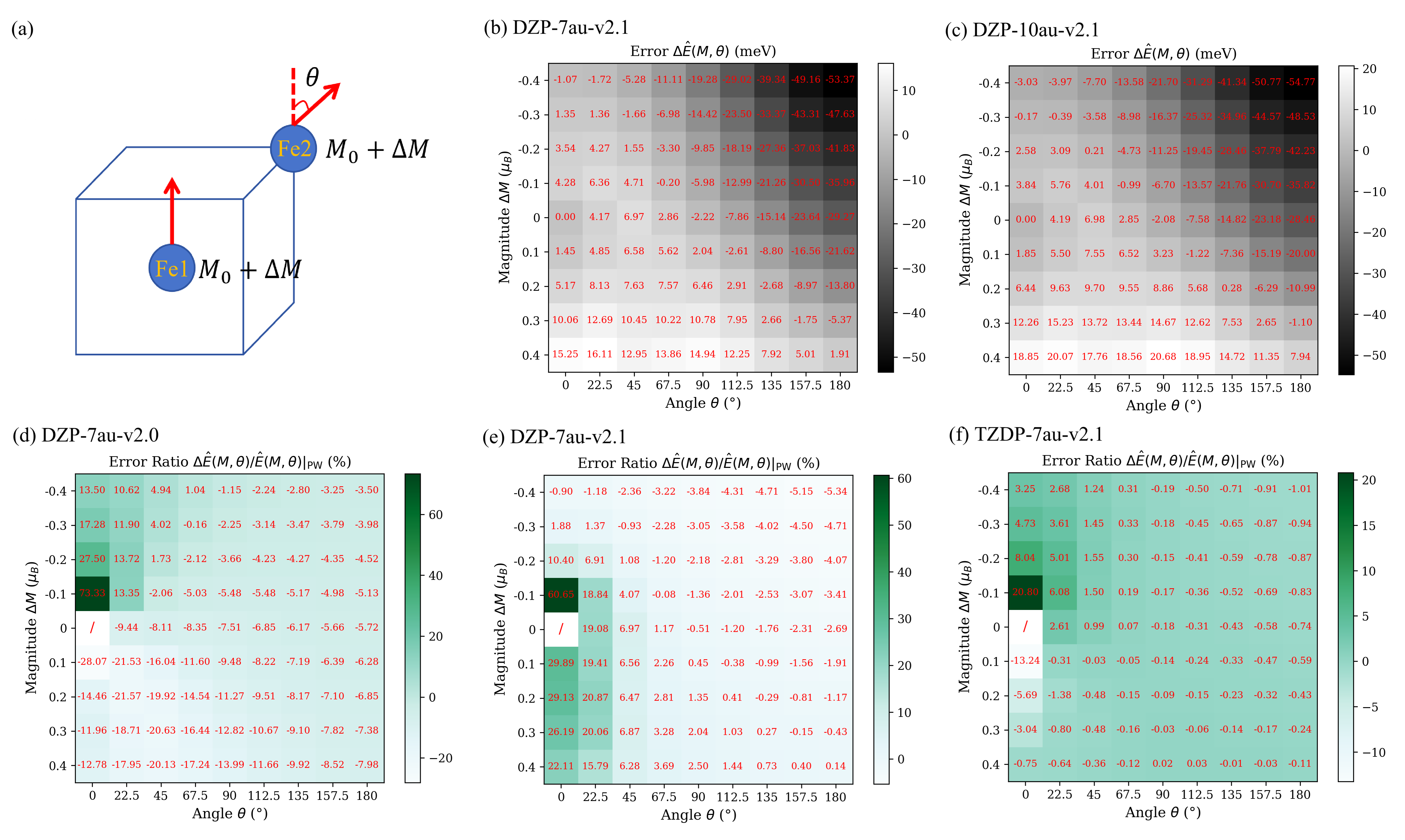}
    \caption{\textbf{The accuracy analysis of different NAO basis.} (a) the BCC-Fe with 2 Fe atoms labels as {$\{M_0-\Delta M,\theta\}$}, where the two atoms have the same atomic magnetic moment $M_0-\Delta M$. $M_0$ is the atomic magnetic moment and $\Delta M$ is ranging from -0.4 $\mu_B$ to 0.4 $\mu_B$. $\theta$ is the angle between them. (b-c) shows the magnetic energy barrier difference between the NAOs basis and PW basis, namely $\Delta\hat{E}(M,\theta)=\hat{E}(M,\theta)|_{\rm NAOs}-\hat{E}(M,\theta)|_{\rm PW}$. Here the magnetic energy barrier is defined by the energy above the ground state $\hat{E}(M,\theta)=E(M,\theta)-E(M_{0},0)$. (d-f) shows the ratio of the magnetic energy barrier difference $\Delta\hat{E}(M,\theta)$ to the PW results.}
    \label{fig:nao_vs_pw}
\end{figure*}

To systematically present the potential energy surfaces of Fe, we modeled BCC-Fe and FCC-Fe with atomic volumes ranging from 8 to 16 \AA$^3$ and calculated their total energies at various magnetic moments using spin-cDFT. The magnetic moments considered ranged from 0.2 to 3.8 $\mu_B$. Fig.~\ref{fig:Fe_2D} illustrates the potential energy surfaces, $E(V,M)$, for BCC-Fe and FCC-Fe, respectively. The results reveal that the BCC phase exhibits a single energy minimum, whereas the FCC phase features two distinct energy minima, consistent with the findings presented in Fig.~\ref{fig:Fe_1D}. In addition, Fig.~S5 displays the potential energy surface $E(\theta,|\mathbf{M}|)$ of BCC-Fe, encompassing both transverse and longitudinal excitations. The results demonstrate that as the angle between the magnetic moments of two neighboring iron atoms increases, the total energy rises progressively, while the magnetic moment amplitude corresponding to the energy-favored configuration gradually diminishes.

Magnetic calculations have high accuracy requirements. Unlike the PW basis, where accuracy can be systematically improved by increasing the number of plane waves, the accuracy of the NAOs basis depends on the quality of the orbitals themselves and lacks a systematic way to enhance precision. In the following, we will quantitatively examine the precision of spin-cDFT results using different basis sets. In the Heisenberg model $H = \sum_{ij}J_{ij}\mathbf{S}_i\cdot\mathbf{S}_j$, the magnetic exchange strength $J$ determines the resistance of the material's magnetic order to external perturbations such as temperature and magnetic fields. This strength typically falls within the range of a few to tens of meV~\cite{2023-misha}. The value of $J$ can generally be derived from the energy differences between various magnetic configurations, making the magnetic barrier energy an excellent quantitative metric for evaluating accuracy.

To this end, we chose a BCC-Fe unitcell containing two Fe atoms, each with the same magnetic moment $M$ and an angle $\theta$ between them. When $\theta=0^{\circ}$, the system represents a ferromagnetic state, and when $\theta=180^{\circ}$, it represents an antiferromagnetic state. To evaluate the quality of orbitals, we selected different NAO basis sets with various cutoff radius $r_c$ and calculated $E(M, \theta)$ for these configurations. Additionally, we used results from a PW basis as a reference. All $E(M, \theta)$ values for these magnetic configurations were computed using spin-cDFT method. It is important to note that the energy errors calculated using the NAOs basis cannot be directly compared to those of the PW basis. Since the magnetic exchange $J$ reflects the response to changes in magnetic moments, we must first subtract the ground-state energy. Specifically, we define  $\hat{E}(M, \theta) = E(M, \theta) - E(M_0, 0)$, where $M_0$ is the atomic magnetic moment of the ground state. We then compare the error in $\hat{E}(M, \theta)$ between the NAOs and PW basis. 

Fig.~\ref{fig:nao_vs_pw} presents the results for various NAOs basis sets, showing the energy error compared to the PW results as $\Delta\hat{E}(M,\theta)=\hat{E}(M,\theta)|_{\rm NAOs}-\hat{E}(M,\theta)|_{\rm PW}$. According to the figure, except in regions where the magnetic moment decreases and the angle is relatively large—where the energy error is significantly higher—the error is fairly uniform in other areas. The results indicate that the error primarily depends on the type of basis rather than the cutoff radius. In practical scenarios, such as the gradual transition from a ferromagnetic to a nonmagnetic state with increasing temperature in BCC-Fe, the magnetic moments significantly change in angle instead of amplitude. When the angle changes from $0^{\circ}$ to $180^{\circ}$, the error for the DZP basis set ranges from 2 to 5 meV, whereas the error for the TZDP basis set is significantly reduced to 0.5 to 1 meV.

To further validate the accuracy within the framework of density functional theory, we analytically calculated the magnetic exchange energy using the magnetic force theorem~\cite{1984-liechtenstein,1987-liechtenstein}. Specifically, we utilized the exchange energy calculations by ABACUS and TB2J~\cite{he2021tb2j}, employing a $9\times9\times9$ supercell to determine the $J_{ij}$ values up to the fourth-nearest neighbors. The results obtained using different NAOs basis are summarized in Table~\ref{tab:JNN}. Since ABACUS currently does not support an interface with PW and TB2J, we used PW results reported in the literature as reference values~\cite{2010-Fe-JNN,2001-Fe-JNN}. Overall, the dominant ferromagnetic $J_{NN}$ calculated by the TZDP basis shows good agreement with the results reported in Ref.~\cite{2001-Fe-JNN}.

\begin{table}[ht]
\caption{\label{tab:JNN} 
\textbf{Exchange energy of ferromagnetic BCC-Fe calculated by various NAOs basis.} The cutoff radii is 7 Bohr for all orbitals. ``NN'' represents the nearest-neighbor exchange, ``2NN'' refers to the second one, and beyond. Ref.~\cite{2010-Fe-JNN} utilizes a private code based on the linear muffin-tin orbital (LMTO) method. Brillouin-zone integrals are calculated by using 1661 irreducible k points. This table shows the result calculated by the PBE functionals. In Ref.~\cite{2001-Fe-JNN}, $J$ is evaluated in the framework of the first principles tight-binding linear muffin-tin orbital method with Vosko-Wilk-Nusair functional. The integration over the full Brillouin zone was performed for very distant coordination shells up to the 195th shell for the bcc lattice.}
\centering
\begin{tabular}{lccccr}
\hline
$J_{ij}$ (meV) & DZP-v2.0 & DZP-v2.1 & TZDP-v2.1 & Ref.~\cite{2010-Fe-JNN} & Ref.~\cite{2001-Fe-JNN}  \\
\hline
NN   & 11.73 & 17.88 & 20.23 & 16.57 & 19.48\\
2NN  & 12.06 & 11.58 & 12.23 & 14.69 & 11.09\\
3NN  & -0.05 & -0.36 & -1.11 & -0.57 & -0.20\\
4NN  & -2.17 & -1.15 & -1.93 & -2.52 & -1.71\\
\hline
\end{tabular}
\end{table}

\subsection{DeePSPIN Model} \label{sec:deepspin}
The non-collinear spin-cDFT implementation in ABACUS provides a powerful tool for studying complex magnetic configurations at the atomic scale. However, conventional DFT methods are time consuming, which makes it challenging to simulate magnetic dynamic processes, such as transitions in magnetic ordering that require large-scale supercells. DeePSPIN~\cite{2024-deepspin} is a deep learning approach for magnetic materials that treats spin as so-called ``pseudo-atoms'' and integrates with the descriptor framework of DeepPot-SE~\cite{18NIPS-Zhang}, preserving translational, rotational, and permutation symmetries. In detail, the DeePSPIN model requires high-precision first-principles data for magnetic materials, including energy, forces, magnetic torques, stress (optional), and atomic configurations, as training data. Additionally, the design of appropriate loss functions and active learning methods~\cite{dpgen} significantly reduces the required number of samples. A well-trained DeePSPIN model can accurately predict physical properties such as energy, forces, and magnetic torques. By combining it with methods like molecular dynamics (MD)~\cite{2022-lammps} and the Landau-Lifshitz-Gilbert equations~\cite{2004-LLG}, it can simulate spin evolution in large-scale systems and accurately describe spin-lattice interactions.

\begin{figure}
    \centering    
    \includegraphics[width=0.7\linewidth]{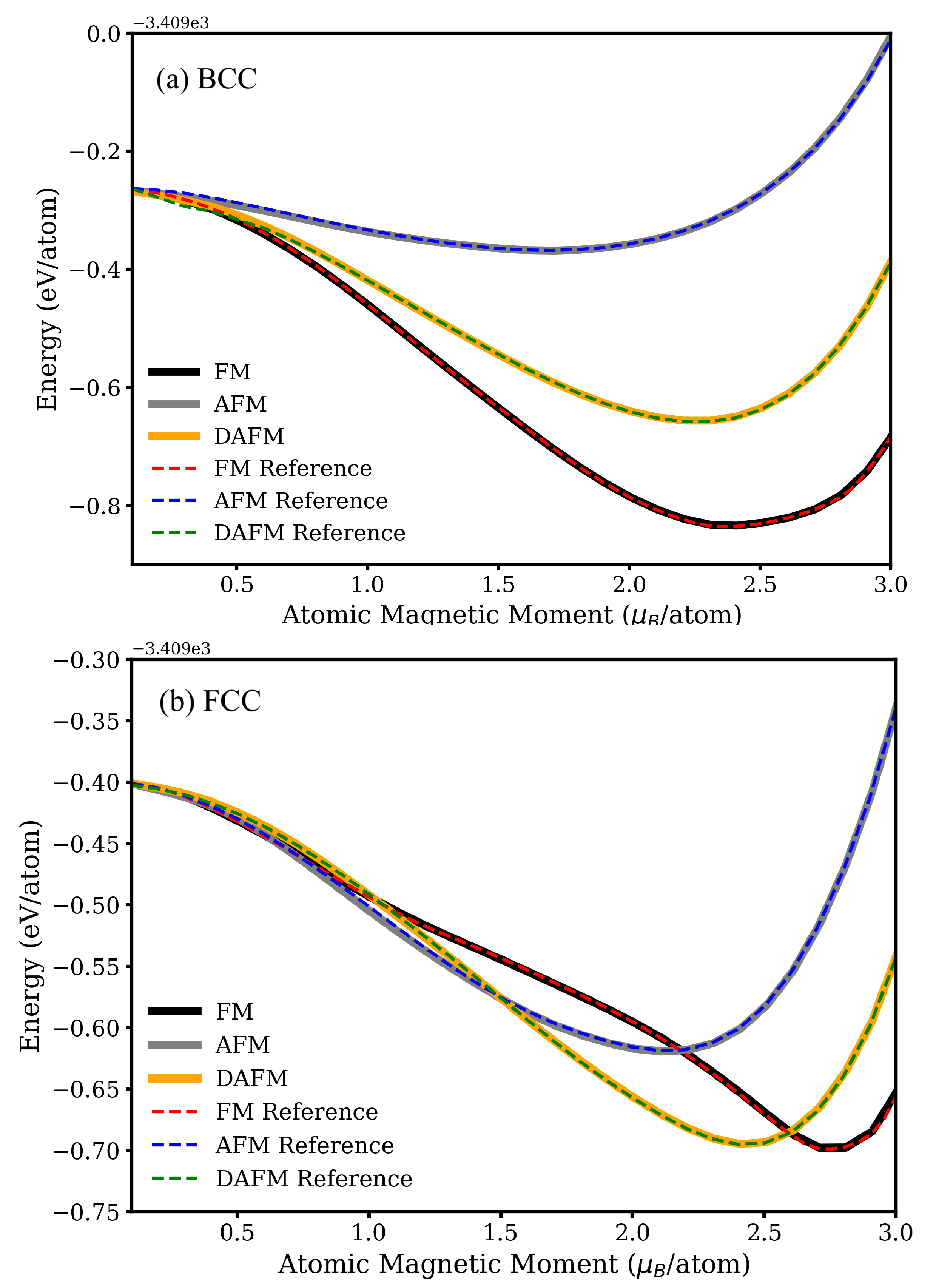}
    \caption{\textbf{The energy comparison between model prediction and DFT results.} (a) For BCC-Fe, the comparison of the energy predicted by the DeePSPIN model with those calculated by ABACUS cDFT for the collinear magnetic configurations (FM/AFM/DAFM) with various atomic magnetic moments. (b) shows the comparison for FCC-Fe.}
    \label{fig:model-configuration}
\end{figure}

\begin{figure}
    \centering    
    \includegraphics[width=0.7\linewidth]{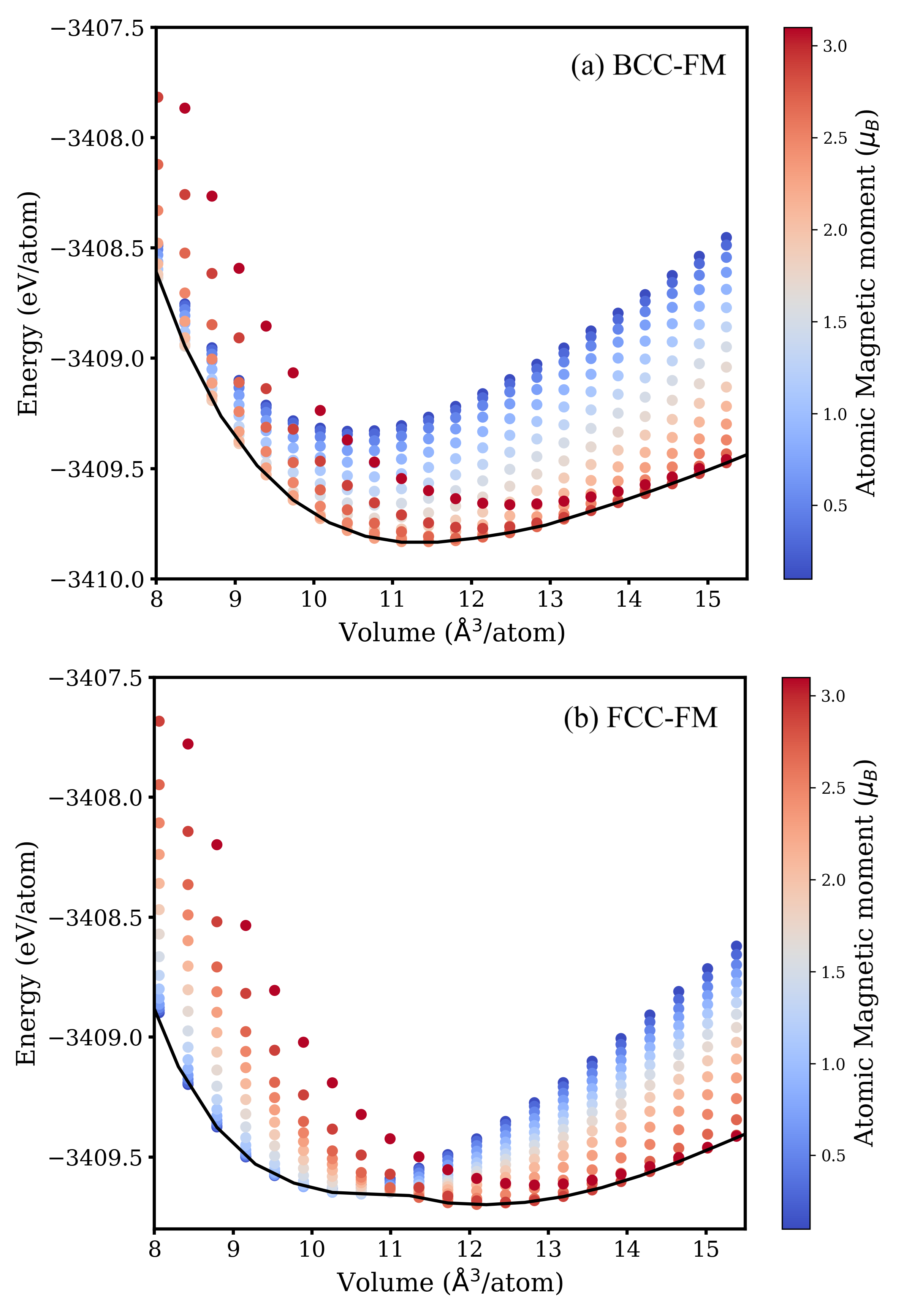}
    \caption{\textbf{The model energy prediction for varying volume.} (a) For FM BCC-Fe and (b) FM FCC-Fe, the energy predicted by the DeePSPIN model as a function of volume with various atomic magnetic moments. The black lines plot the energy calculated by unconstrained DFT method.}
    \label{fig:model-evm}
\end{figure}

To assess the capability of ABACUS+DeePSPIN in representing significant changes in the lattice and magnetic moments, we generated over 10,000 collinear magnetic configurations for the BCC/FCC phase of iron, including three magnetic states of ferromagnetic (FM), antiferromagnetic (AFM), and double-layer ferrimagnetic (DAFM). The cell volume perturbations ranged from -30\% to +30\%, and the magnitude of the atomic magnetic moments varied from 0 to 4.0 $\mu_B$. These data points were obtained through collinear spin-cDFT calculations using the DZP-8au-v2.1 basis set. We then trained a DeePSPIN model based on these perturbed configurations. Fig.~\ref{fig:model-configuration} adopts both DeePSPIN and DFT methods to predict the total energy of collinear magnetic configurations such as FM, AFM, and DAFM. We find the results from the two models align well with each other, demonstrating the good accuracy of the DeePSPIN model. Fig.~\ref{fig:model-evm} illustrates the dependence of total energy for varying magnitudes of magnetic moments, with the volume being changed isotropically. The solid line represents the ground-state energy from unconstrained DFT calculations at each volume, while the data points from the DeePSPIN model correspond to excited states, which have energies higher than those of the DFT results. These observations are in good agreement with recent studies~\cite{2024-fe-npj}, demonstrating the reliability of the ABACUS spin-cDFT data and the expressive capability of the DeePSPIN model.

To explore more complex magnetic phenomena, such as the transition of magnetic ordering, it is crucial for the model to accurately capture information about non-collinear magnetic moments. To this end, we trained the DeePSPIN model following the automated workflow, comprising four key components: (1) initial configuration construction, (2) first-principles data labeling, (3) model training, and (4) sampling exploration. The workflow begins with the construction of an initial dataset. Based on two fundamental structures (BCC and FCC) and three magnetic configurations (FM, AFM, and DAFM), we obtained six initial configurations ($2\times2\times2$ supercells containing 16 Fe atoms). Subsequently, random perturbations were applied to each configuration’s atomic positions, lattice strain, magnetic moment orientations, and magnetic moment magnitudes, generating 100 perturbed configurations per initial structure to form the initial dataset. The maximum perturbation amplitude for the lattice cell was 1\%, while atomic positions were perturbed by up to 0.03 $\rm\AA$. Magnetic moments were randomly rotated by up to 90 degrees, and their magnitudes were perturbed by up to 0.5 $\mu_B$. The initially sampled configurations were subjected to first-principles calculations using ABACUS to obtain corresponding physical quantities such as energy ($E$), atomic forces ($\mathbf{F}$), stress ($\sigma$), and magnetic torque ($\boldsymbol{\lambda}$). These data, combined with random seeds, were used to train four different initial DeePSPIN magnetic models~\cite{2024-deepspin}. Starting from these models, we performed active learning, which consisted of four components: configuration exploration, configuration analysis, first-principles sampling, and model updating.

The exploration was conducted based on an active learning strategy~\cite{dpgen}, employing molecular dynamics~\cite{2025-huang-prep}. The molecular dynamics simulations utilize a modified version of LAMMPS, explicitly adapted to include atomic magnetic moment dynamics and their coupling with lattice motions~\cite{2025-huang-prep}. Specifically, we selected one of the models to perform molecular dynamics simulations and sample configurations, while using all four initial models to predict material properties for the new configurations. The active learning strategy selects new configurations based on the models' prediction uncertainty, specifically identifying ``high-value'' configurations as those exhibiting large prediction uncertainties. Configurations exhibiting smaller errors are categorized as ``well-learned'', while those with anomalously large errors are labeled as ``unphysical''. Both types of configurations are subsequently removed from the exploration space. High-value configurations were then subjected to first-principles calculations. The newly acquired data were incorporated into the existing training set to update the models. The exploration process started at low temperatures, and after model convergence within each temperature interval, the MD simulation temperature was incrementally increased. The sampling temperature ranged from 50 to 1600~K, with each temperature range including MD configurations with the virtual magnetic mass of 0.01/0.05/0.1/0.5/1.0. The virtual magnetic mass serves as an auxiliary parameter introduced to characterize spin dynamics in the simulations. While this parameter influences the numerical stability of the calculations, our tests demonstrate that it does not introduce any significant bias in the magnetic evolution (see Fig.~S10). This result equivalently demonstrates that varying magnitudes of magnetic torque do not affect the thermodynamic statistical outcomes. We employed both NAOs basis (DZP-7au-v2.0) and PW basis to generate two separate data sets. All computations were performed using non-collinear spin-cDFT, incorporating spin-orbit coupling. The PW training strategy followed the same approach as for the NAOs, with the key difference being that NAOs BCC sampling involved 400 configurations per round, whereas PW BCC sampling involved 200 configurations per round. Ultimately, the NAOs basis set generated a total of 34,703 DPGEN samples, while the PW basis set generated 20,931 DPGEN samples. Additionally, for each basis set, we supplemented the data with 175 perturbed configurations for FCC-FM and 109 configurations for the FCC $4 \times 2 \times 2$ supercells. 

\begin{figure}
    \centering    
    \includegraphics[width=0.7\linewidth]{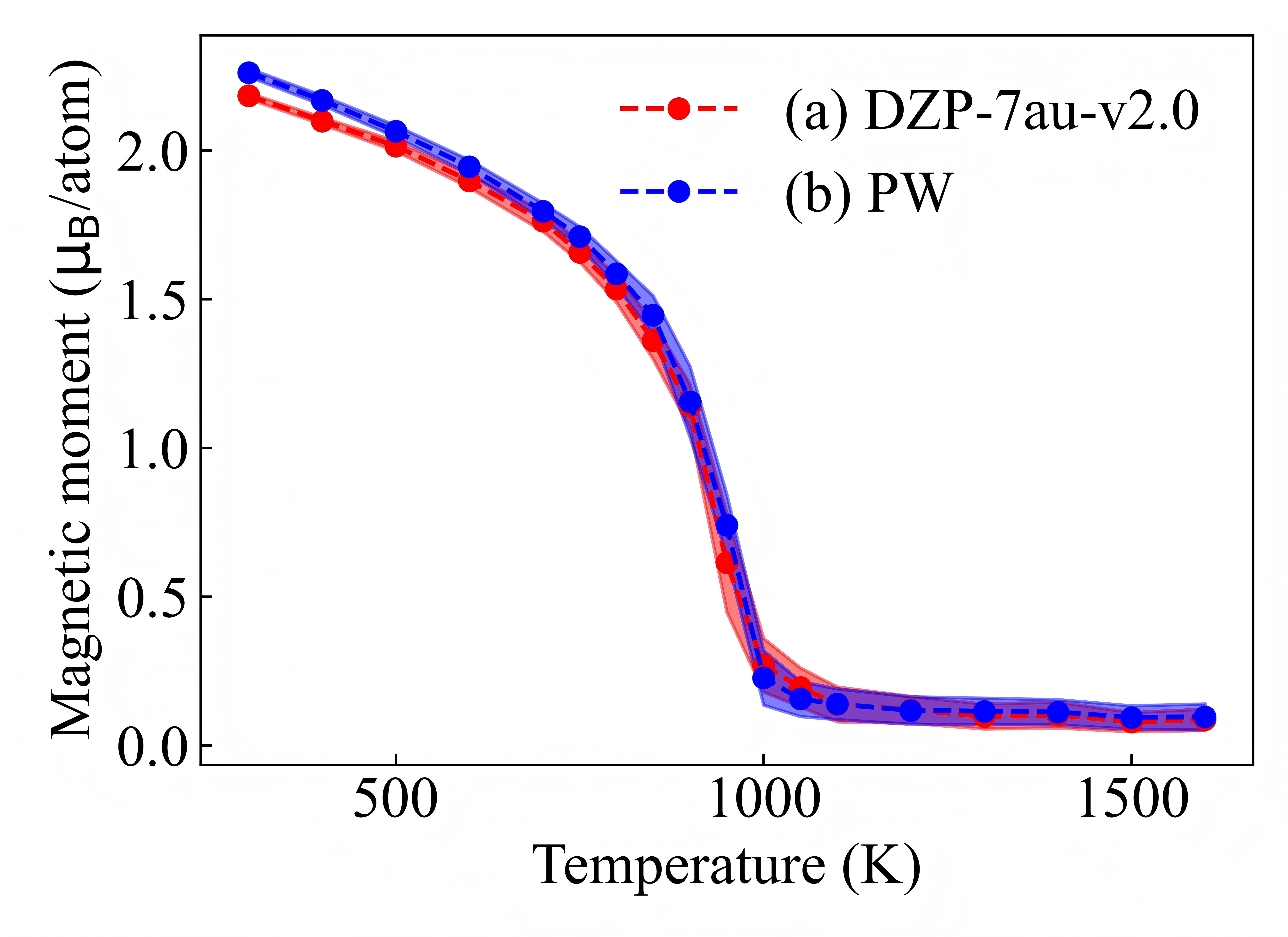}
    \caption{\textbf{The ferromagnetic-paramagnetic transition.} Magnetization of BCC-Fe as a function of temperature. Based on the DeePSPIN model, MD simulations at different temperatures are performed, with the stable magnetic moment counted as the magnetization at that temperature. Line width refers to the error bar. The number of atoms in the supercell is 1024, and the virtual magnetic mass M=0.01.}
    \label{fig:Tc}
\end{figure}

We trained two models, DeePSPIN-DZP and DeePSPIN-PW, based on first-principles data, including energy, force, and magnetic torque. These models allow us to perform large-scale simulations to observe the evolution of magnetism with temperature. The training errors and test performance of the DZP model are presented in the Supplementary Information (see Fig.~S1 and Fig.~S2). While the DZP basis set introduces a systematic energy error of approximately 1 meV/atom, this magnitude of error is well within the acceptable tolerance range for both model training and prediction. The basis set error does not constitute the primary source of model's training error. The model demonstrates consistency in prediction errors across different system sizes as depicted in Fig.~S3. We conducted NVT ensemble simulations at different temperatures using the trained models, employing an $8 \times 8 \times 8$ supercell (consisting of 1024 atoms), with a virtual magnetic mass set to 0.01, a timestep of 0.1~fs, for a total of 3~ps of simulation. In Fig.~S11, we present the time-dependent magnetic moment $\mathbf{M}(t)$ at several temperatures. After the MD equilibration, we selected the trajectory from 1 to 3 ps to calculate the average total magnetization of the system. In Fig.~\ref{fig:Tc}, we present the total magnetization of BCC-Fe as a function of temperature. It can be observed that BCC-Fe exhibits ferromagnetic behavior at low temperatures, with the total magnetization exceeding 2 $\mu_B$. The magnetization decreases gradually up to 800~K, before undergoing a sudden drop around 1000~K, at which point it approaches zero, indicating a ferromagnetic-to-paramagnetic transition (see Fig.~S12). This transition temperature is in close agreement with the experimentally observed Curie temperature of 1043~K. Furthermore, systematic analysis of dynamical simulation convergence with respect to system size demonstrates that while the model overestimates total magnetic moments in small systems at elevated temperatures (due to challenges in modeling paramagnetic states), its predictions exhibit progressive convergence as system dimensions increase, as evidenced in Fig.~S13. Surprisingly, although the DZP-7au-v2.0 basis set introduces non-negligible errors when calculating the magnetic exchange strength $J$, the magnetization as a function of temperature and the Curie temperature is in very close agreement with the results from PW basis.

In summary, we have implemented the ABACUS spin-cDFT method, which combines first-principles data with AI-assisted magnetic models to extend the precision of magnetic first principles to larger scales, which allows molecular dynamics simulations of lattice and magnetic moment behavior at large spatial scales and long-time scales. We demonstrate that the deep learning magnetic model accurately predicts the Curie temperature of iron. Furthermore, the model can be used to study complex dynamic phenomenon such as the BCC-FCC structural phase transition of iron. A more detailed discussion of this will be provided in a forthcoming paper.

\section{Discussion} \label{sec:conclusion}
In this study, we have implemented a non-collinear spin-constrained method within the open-source software ABACUS, utilizing both plane wave and numerical atomic orbital basis. This implementation allows for the precise control and calculation of arbitrary magnetic configurations. We introduce a smooth modulation orbital method for calculating atomic magnetization using the NAOs projection. Systematic investigations on bulk iron have demonstrated its reliability. The precision of the numerical atomic orbitals used in magnetic calculations has been quantitatively discussed, with the TZDP basis set yielding results very close to the reference plane-wave basis.

An automated workflow is utilized to train a fundamental magnetic DeePSPIN model for elemental Fe. Combined with molecular dynamics simulations, the DeePSPIN model based on PW and NAO basis both successfully observed the ferromagnetic-paramagnetic transition near the experimental Curie temperature, demonstrating the robustness and effectiveness of this workflow. 
The datasets and models presented in this paper will be made openly available to support subsequent fine-tuning and applications. The dataset in this study was constructed from the outset based on fully constrained cDFT. This conservative approach may lead to redundancy in the dataset. The distribution of magnetic moment magnitudes remains relatively concentrated in the low-temperature region. A feasible solution is to initially employ direction-only constrained cDFT for sampling in the low-temperature region, while gradually increasing the proportion of fully constrained cDFT calculations for magnetic moment magnitude variations as the temperature rises. This hybrid strategy is expected to significantly reduce the required number of samples while maintaining model accuracy. We anticipate that future work will address this limitation.

In conclusion, the noncollinear spin-constrained method implemented in ABACUS provides a powerful tool for studying complex magnetic phenomena at the atomic scale, and a data engine for deep-learning magnetic models. By integrating first-principles calculations, dynamical simulations, active learning, and magnetic modeling methods, we transfer first-principles accuracy to large-scale magnetic simulations in a maximally automated way. This integration enables large-scale magnetic simulations, providing new possibilities for the study of complex magnetic phenomena.

\section{Method} \label{sec:method}

\subsection{Non-Collinear Spin}
For systems with non-collinear spin configurations, where the spin is not uniformly aligned along a single axis, the electronic wavefunction is expressed as a spinor $\psi = \{\psi^{\uparrow}, \psi^{\downarrow}\}$. In the context of non-collinear magnetism within density functional theory, the Kohn-Sham equations for a two-component spinor wave function $\psi$ can be written as~\cite{1965-ks}
\begin{equation}
 \left[ -\frac{\hbar^2}{2m}\nabla^2 + V_{\text{eff}}(\mathbf{r}) \right]\psi_i(\mathbf{r}) = \epsilon_i \psi_i(\mathbf{r}),
\end{equation}
where $V_{\text{eff}}$ is the effective potential, including the Hartree term $V_{\text{H}}$, the external potential term $V_{\text{ext}}$ and the exchange-correlation (XC) term $V_{\text{xc}}$. In addition, the generalized charge density can be defined by introducing two-component spin space and the formula is
\begin{equation}
\bar{\rho}=\frac{1}{2}(\rho + \sum_{p}m_p\cdot\boldsymbol\sigma^p)=\frac{1}{2}\left(\begin{array}{cc} \rho + m_{z} & m_{x}-i m_{y} \\ m_{x}+i m_{y} & \rho-m_{z} \end{array}\right),
\end{equation}
where
\begin{equation}
\rho(\mathbf{r})=\sum_{i}f_i\psi_{i}^{\dagger}(\mathbf{r})\psi_{i}(\mathbf{r})
\end{equation}
is charge density and the spin density vector $\mathbf{m}=(m_{x}, m_{y}, m_{z})$ is
\begin{equation}
m_p(\mathbf{r})=\sum_{i}f_i\psi_i^{\dagger}(\mathbf{r})\boldsymbol\sigma^p\psi_i(\mathbf{r}).
\end{equation}
Here $\psi_i$ is two component spinor wavefunction for $i$-th band with occupation number $f_i$. $p$ takes values \{1, 2, 3\}, representing the components of spin density $\mathbf{m}$ along the $x$, $y$ and $z$ directions. At the same time, it also denotes the $p$-th Pauli matrix $\boldsymbol\sigma^p$.

The magnetic effects in the Hamiltonian can be categorized into two parts. First, the electron-electron interaction, which is encompassed within the exchange-correlation term, directly influences the self-consistent solution process and necessitates the careful selection of appropriate functionals. Second, the spin-orbit coupling (SOC) effect, arising from relativistic effects, is introduced through fully relativistic pseudopotentials and is computed within the non-local components of the pseudopotential.

In the computation of the XC functional, the treatment of non-collinear spins can be simplified by using the local density approximation (LDA), which reduces the problem to calculating collinear spins at each grid point. For local reference systems used to calculate the XC functional, the local charge density can be transformed into the diagonal matrix form of the generalized charge density as
\begin{equation}
\bar{\rho}^{\prime}=\left(\begin{array}{cc} \rho_{+} & 0 \\ 0 & \rho_{-} \end{array}\right),
\end{equation}
where $\rho_+$ and $\rho_-$ are defined as 
\begin{equation}
\rho_{\pm}=\frac{1}{2}(\rho \pm |\mathbf{m}|)=\frac{1}{2}\left(\rho \pm\sqrt{m_{x}^{2}+m_{y}^{2}+m_{z}^{2}}\right).
\end{equation}

Taking into account the non-collinear electron spins within the framework of the aforementioned local spin density approximation (LSDA)~\cite{1989-lsda}, the expressions for the exchange-correlation energy functional takes the form of
\begin{equation}
 E_{\mathrm{XC}}(\rho(\mathbf{r}),\mathbf{m}(\mathbf{r}))=\int
\rho(\mathbf{r}) \varepsilon_{\mathrm{XC}}(\rho(\mathbf{r}),|\mathbf{m}(\mathbf{r})|)d\mathbf{r},
\end{equation}
and the corresponding potential is
\begin{equation}
\begin{aligned}
 V_{\mathrm{XC}}(\mathbf{r})&=\frac {\delta E_{\mathrm{XC}}}{\delta \rho(\mathbf{r})}\\ &= \varepsilon_{\mathrm{XC}} (\rho(\mathbf{r}),|\mathbf{m}(\mathbf{r})|)+\rho(\mathbf{r}) 
 \left[ \frac {|\partial \varepsilon_{\mathrm{XC}}(\rho(\mathbf{r}),|\mathbf{m}(\mathbf{r})|)}{\partial \rho(\mathbf{r})} \right].
\end{aligned}
\end{equation}
The computation of the XC magnetic field $\mathbf{b}(\mathbf{r})$ can be achieved by employing the chain rule to differentiate the modulus of the spin density, ensuring that $\mathbf{b}(\mathbf{r})$ is collinear with the magnetization vector $\mathbf{m}(\mathbf{r})$. The formula is
\begin{equation}
\begin{aligned}
\mathbf{b}(\mathbf{r})&=-\frac {\delta E_{\mathrm{XC}}}{\delta \mathbf{m}(\mathbf{r})} = -\frac {\delta |\mathbf{m}(\mathbf{r})|}{\delta \mathbf{m}(\mathbf{r})} \frac {\delta E_{\mathrm{XC}}}{\delta |\mathbf{m}(\mathbf{r})|} \\
&=- \widehat {\mathbf{m}}(\mathbf{r}) \rho(\mathbf{r}) \left[ \frac {\partial \varepsilon_{\mathrm{XC}}(\rho(\mathbf{r}),|\mathbf{m}(\mathbf{r})|)}{\partial |\mathbf{m}(\mathbf{r})|} \right].
\end{aligned}
\end{equation}

The generalized gradient approximation (GGA) functionals~\cite{perdew1996generalized}, which incorporate both charge density and its gradient as well as the spin density gradient, are widely used because of the excellent balance between accuracy and efficiency. For non-collinear spin calculation, the gradient of the spin density vector can be calculated in different ways~\cite{kubler1988density, sjostedt2002noncollinear, scalmani2012new, peralta2007noncollinear}, while none of them simultaneously achieves both high efficiency and avoidance of the well-known numerical instabilities~\cite{peralta2007noncollinear, scalmani2012new} for non-collinear-spin GGA functionals. The method implemented in ABACUS, proposed by K$\ddot{u}$bler \textit{et al.}~\cite{kubler1988density, sjostedt2002noncollinear}, calculates $\nabla \rho_{\pm}(\mathbf{r})$ without considering the different rotation matrices that would reduce non-collinear spins to collinear spins at various grids, treating $\rho_{\pm}(\mathbf{r})$ as a scalar function.

The spin-orbit coupling and relativistic effects are incorporated using norm-conserving pseudopotentials~\cite{hamann1979norm,troullier1991efficient} within the Kleinman-Bylander (KB) form ~\cite{hemstreet1993first, theurich2001self,kleinman1980relativistic, bachelet1982relativistic}.
Additionally, spin-orbit terms are directly included in both plane-wave basis sets~\cite{corso2005spin} and numerical atomic orbitals basis sets~\cite{cuadrado2012fully} for self-consistent field calculations.

\subsection{Projection Methods}\label{sec:proj_method}
As the core physical quantity in spin-cDFT, the atomic magnetic moment is crucial for accurate algorithmic implementation. Typically, one can define atomic magnetic moments in three ways: by partitioning the spin density \cite{bader1990quantum, fonseca2004voronoi, hirshfeld1977bonded}, through wavefunction analysis \cite{mulliken1955electronic, lowdin1970nonorthogonality}, and via subspace projection \cite{hegde2020atomic}.

A straight definition belonging to the first kind is the ``spherical definition method'', in which the spin density is integrated over a spherical volume around the nucleus that is defined by the window function $\Theta(\mathbf{r};\boldsymbol\tau_I)$. For example 
\begin{equation}
\mathbf{M}_I = \int \mathbf{M}(\mathbf{r}) \Theta(r^{\text{cut}} - |\mathbf{r} - \boldsymbol\tau_I|)\mathrm{d}\mathbf{r},
\end{equation}
where $\boldsymbol\tau_I$ denotes the position of the nucleus $I$ and $r^{\mathrm{cut}}$ is the cutoff radius, $\Theta$ is the Heaviside step function or other smoothed window function. 

More sophisticated strategy requires the conservation on quantities to be partitioned, 
\begin{equation}
    \mathbf{M}_I\left( \mathbf{r} \right) =\mathbf{M}\left( \mathbf{r} \right) w_I\left( \mathbf{r} \right), 
\end{equation}\label{eq:density-partition}
in which
\begin{equation}
    w_I\left( \mathbf{r} \right) =\frac{P_I\left( \mathbf{r} \right)}{\sum_J{P_J\left( \mathbf{r} \right)}}.
\end{equation}
$w_I(\mathbf{r})$ is the weighting function and partition function $P_I(\mathbf{r})$ approaches to 1 near $I$-th atom and 0 otherwise. A famous example of a more sophisticated partitioning strategy is called Atom-In-Molecule (AIM), or well-known Bader charge analysis~\cite{bader1990quantum}, which defines atomic regions by zero-flux surfaces in the gradient of the electron density, Voronoi tessellation~\cite{fonseca2004voronoi} partitions space into polyhedral cells around each atom based on proximity. The Bader charge analysis provides a mathematically rigorous definition of atomic boundaries based on electron density, yielding charges that are invariant to the choice of basis set and more reflective of the true electronic structure. However, this method would fail in cases where there are zero or more than one atoms appear in one Voronoi cell, and lack of decomposition of the orbits.

Mulliken~\cite{mulliken1955electronic} and Löwdin~\cite{lowdin1970nonorthogonality} charge analysis methods are well-established for partitioning electron density within a molecule to assign atomic charges. The formula for calculating the atomic magnetization from Mulliken charge of $I$-th atom is given by
\begin{equation}
\mathbf{M}_I^{\mathrm{Mulliken}} = \sum_{\mu\in I}\sum_{\mathbf{k}}\sum_{\nu}\bm{\sigma}\cdot\bm{\rho}_{\mu\nu}(\mathbf{k})S_{\mu\nu}(\mathbf{k}).
\end{equation}
Similarly, the formula for the Löwdin charge is:
\begin{equation}
\mathbf{M}_I^{\mathrm{L\ddot{o}wdin}} = \sum_{\mu\in I}\sum_{\mathbf{k}}\sum_{\nu}\bm{\sigma}\cdot S_{ \mu\nu}^{1/2}(\mathbf{k})\bm{\rho}_{\mu\nu}(\mathbf{k})S_{\mu\nu}^{1/2}(\mathbf{k}),
\end{equation}
where $\bm{\sigma}$ is the Pauli matrix to decompose reciprocal space density matrix $\bm{\rho}_{\mu\nu}(\mathbf{k})$, and $S_{\mu\nu}(\mathbf{k})$ is the overlap integral matrix between basis functions $\mu$ and $\nu$ on this atom. These two charge analysis methods are powerful tools for approximating electron distribution by assigning shared electrons between atoms based on orbital overlaps, thereby offering a rudimentary yet quick insight into molecular charge states. However, their reliance on the choice of basis sets and the arbitrary nature of electron partitioning, which overlooks electronegativity differences, often results in charges that lack physical accuracy and can vary significantly with computational parameters.

The subspace projection method requires a rational construction of projection operators in which the projection function should be of physical meaning. For example in projected augmented wave (PAW) formulism~\cite{1994-paw}, the PAW projector $\{|p_{I\mu}\rangle\}$ which holds the orthonormality with the pseudo partial wave that is connected with the all-electron one $\{|\phi_\mu\rangle\}$ is a natural choice~\cite{2015-ma,2020-SPHInX}. Atomic magnetic moments is defined as
\begin{equation}
    \mathbf{M}^p_I = \sum_{\sigma\sigma'}\sum_{\mu\nu}\bm{\sigma}^{p\sigma\sigma'}D_{I\mu\nu}^{\sigma\sigma'}\Omega_{\mu\nu},
\end{equation}
in which the scalar $\sigma$ and $\sigma'$ refer to the spin index (up and down) and the $\bm{\sigma}^p$ is the $p$-th Pauli matrix. $D_{I\mu\nu}^{\sigma\sigma'}$ is the one-center density matrix representation of pseudo partial wave centered at the $I$-th atom, 
\begin{equation}
    D_{I\mu\nu}^{\sigma\sigma'} = \sum_n f_{n}^\sigma\langle\psi_{n}^\sigma|p_{I\mu}\rangle\langle p_{I\nu}|\psi_{n}^{\sigma^\prime} \rangle,
\end{equation}
$\Omega_{\mu\nu}$ is the all-electron partial wave matrix elements of the cutoff-sphere integral 
\begin{equation}
    \Omega_{\mu\nu} = \int d^3\mathbf{r}\phi_\mu(\mathbf{r})\phi_\nu(\mathbf{r})\Theta(r^\mathrm{cut}-|\mathbf{r} - \mathbf{\boldsymbol\tau}_I|).
\end{equation}
This algorithm effectively utilizes the PAW projection orbitals’ ability to accurately describe different electronic angular momentum orbitals, while also leveraging spherical truncation to preserve locality, making it a highly efficient method for defining atomic magnetic moments.

In this work, we propose a projection method under pseudopotential formalism for calculating atomic magnetization,
we employ valence electron orbital local density projection operators 
\begin{equation}
\hat{P}_{Ilmm'} = | \alpha_{Ilm}\rangle\langle \alpha_{Ilm'}| 
\end{equation}
to estimate atomic magnetic moments $\mathbf{M}_I$, where $\alpha_{Ilm}$ is the projection orbital distinguished by the angular momentum quantum number $l$ and the magnetic quantum number $m$. Within this framework, the sum of occupancy of the valence electrons $\sum_{m}n^{\sigma\sigma'}_{Ilm}$ with identical angular momentum $l$ and spin index $\sigma,\sigma'$ corresponds to the trace of the projected density matrix $ \text{Tr}(n^{\sigma\sigma'}_{Ilmm'})$, where 
\begin{equation}
n^{\sigma\sigma'}_{Ilmm'} = \sum_{i}f_i\langle\psi^\sigma_i|\hat{P}_{Ilmm'}|\psi^{\sigma'}_i\rangle. 
\end{equation}
The atomic magnetic moment along the direction $p$ is calculated by summing the diagonal elements of the occupation matrix for each angular momentum
\begin{equation}
    \mathbf{M}_{I}^{p}=\sum_{\sigma \sigma ^{\prime}}{{\sum_{lmm^{\prime}}{\bm{\sigma}_{\sigma \sigma^{\prime}}^{p}n_{Ilmm^{\prime}}^{\sigma \sigma^{\prime}}\delta _{mm^{\prime}}}}},
\label{atomic_mag}
\end{equation}
where $p$ refers to $x, y, z$ components.

The accuracy and locality of the projection depend upon the orbitals $|\alpha_{Ilm}\rangle$ used to construct the projection operators $\hat{P}_{Ilmm'}$, which are composed of radial distribution functions and spherical harmonics 
\begin{equation}
\alpha_{Ilm}(\mathbf{r}) = \alpha_{l}(|\mathbf{r}-\bm{\tau}_I|)Y_{lm},
\end{equation}
where the radial functions $\alpha(r)$ are constructed to meet three criteria: (1) maximized efficiency for extracting valence electron information of interest, (2) normalization, and (3) smoothness at the boundary to avoid numerical error. Practically, we construct smooth modulation orbitals (SMOs) by truncating those $\zeta$ functions of ABACUS numerical atomic orbitals (NAOs)~\cite{lin2021strategy, Chen2010} that collect majority of valence electron information over various reference systems, then smoothed by a function centered at the boundary

\begin{equation}
    g(r;\sigma)=\left\{\begin{matrix}
1-\exp\left(-\frac{(r-r_m)^2}{2\sigma^2}\right) & r < r_m \\
0&r\geq{}r_m
\end{matrix}\right. ,
\label{eq:orb_mod}
\end{equation}
where $r$ is the distance to the atom, $r_m$ is the artificially chosen modulation radius. Spreading parameter $\sigma$ is solved in an iterative way, that can minimize the gradient term of spillage~\cite{lin_accuracy_2020} between the SMOs  $|\alpha(\sigma)\rangle$ and NAOs $|\chi\rangle$ under the constraint of normalization of the SMOs themselves
\begin{equation}
\min_{\sigma \in \mathbb{R} ^+} \left\| \nabla |\chi \rangle -\nabla |\alpha \left( \sigma \right) \rangle \right\| ^2\,\,\mathrm{s}.\mathrm{t}.
\langle \alpha \left( \sigma \right) |\alpha \left( \sigma \right) \rangle =1,
\end{equation}
where
\begin{equation}
\langle \mathbf{r}|\alpha _{lm}\left( \sigma \right) \rangle =\chi _{lm}\left( r \right) g\left( r;\sigma \right) Y_{lm}\left( \hat{\mathbf{r}} \right).
\end{equation}
Using SMOs as projection orbitals allows for the representation of charge distributions across different angular momentum orbitals within an atom’s localized environment. Since the projection orbitals are designed to be strongly localized, inter-atomic overlap can be neglected. Furthermore, the orthogonality of projection orbitals with different angular momenta is preserved during modulation, resulting in a simplified form for the projection operator.

The onsite occupation matrix, which quantifies the localized atomic charge occupancy, can be obtained through the projection of wave functions onto SMOs. In NAO formulation for periodic systems, the onsite occupation matrix can be written as
\begin{equation}
\begin{aligned}
    n^{\sigma\sigma'}_{Ilmm'} 
    &= \sum_{n\mathbf{k}} f_{n\mathbf{k}} \langle\psi_{n\mathbf{k}}^\sigma |\alpha_{Ilm}\rangle\langle\alpha_{Ilm'}|\psi_{n\mathbf{k}}^{\sigma^\prime}\rangle \\
    &= \sum_\mathbf{R}\sum_{\mu\nu}\rho_{\mu\nu}^{\sigma\sigma'}(\mathbf{R})\langle\phi^{\mathbf{0}}_{\mu} |\alpha_{Ilm}\rangle\langle\alpha_{Ilm'}|\phi^{\mathbf{R}}_{\nu}\rangle,
\end{aligned}
\end{equation}
where 
\begin{equation}
|\psi_{n\mathbf{k}}^\sigma\rangle=\sum_{\mu}c^{\sigma}_{n\mu}(\mathbf{k})\phi_{\mu}(\mathbf{k})
\end{equation}
represents the eigenvector of $n$-th band at the first Brillouin zone sampling point $\mathbf{k}$, and the spin index is $\sigma$. The parameter $f_{n\mathbf{k}}$ depicts occupation numbers of electrons.
$\phi^{\mathbf{0}}_{\mu}=\phi_\mu(\mathbf{r}-\boldsymbol\tau_{\mu})$ and $\phi^{\mathbf{R}}_{\nu}=\phi_\nu(\mathbf{r}-\mathbf{R}-\boldsymbol\tau_\nu)$ are numerical orbital functions. $\rho_{\mu\nu}^{\sigma\sigma'}(\mathbf{R})$ is the real space density matrix that takes the form of
\begin{equation}
\rho_{\mu\nu}^{\sigma\sigma'}(\mathbf{R}) = \frac{1}{N_\mathbf{k}}\sum_{n\mathbf{k}}f_{n\mathbf{k}}c_{n\mu}^{\sigma *}(\mathbf{k})c_{n\nu}^{\sigma'}(\mathbf{k})e^{i\mathbf{k\cdot R}}.
\end{equation}

For plane wave basis, the onsite occupation matrix $n^{\sigma\sigma'}_{Ilmm'}$ can be written as
\begin{equation}
\begin{aligned}
    n^{\sigma\sigma'}_{Ilmm'} 
    &= \sum_{n\mathbf{k}} f_{n\mathbf{k}} S_{In\mathbf{k}lm}^{\sigma*}S_{In\mathbf{k}lm'}^{\sigma^\prime},
\end{aligned}
\label{occ_matrix}
\end{equation}
where
\begin{equation}
S_{In\mathbf{k}lm}^{\sigma}=\sum_\mathbf{G}{\alpha_{Ilm}^{*}(\mathbf{G})c_{n\mathbf{k}}^\sigma(\mathbf{G})}
\end{equation}
is the plane wave basis representation of the overlap between $|\psi_{n\mathbf{k}}^\sigma\rangle$ and SMOs, $c_{n\mathbf{k}}^\sigma(\mathbf{G})$ and $\alpha_{Ilm}(\mathbf{G})$ are coefficients of plane wave expansion of $|\psi_{n\mathbf{k}}^\sigma\rangle$ and SMOs, respectively. One can refer to Ref.~\cite{king1991real} for the details of SMOs expansion.

\subsection{\label{sec:spin-cDFT}Spin-Constrained DFT}
Utilizing the established Lagrange formalism~\cite{dederichs1984ground, 2023-cai}, the challenge pertaining to the restriction of atomic magnetic moments is reformulated as an endeavor to ascertain the equilibrium points of the pertinent function. To this end, a Lagrange multiplier is introduced into the energy functional of the system:
\begin{equation}
\begin{aligned}
E_c = E_{\text{KS}}&\big(\rho(\mathbf{r}), \mathbf{m}(\mathbf{r})\big) \\
&- \sum_I \boldsymbol{\lambda}_I \cdot \Big(\mathbf{M}_I\big(\mathbf{m}(\mathbf{r})\big) - \mathbf{M}_{I,\text{target}}\Big) ,
\end{aligned}
\label{ec_deltaspin}
\end{equation}
where $ E_{\text{KS}} $ is the Kohn-Sham energy with charge density $\rho(\mathbf{r})$ and spin density $\mathbf{m}(\mathbf{r})$, $\boldsymbol{\lambda}_I$ is the Lagrange multiplier, which can be treated as the magnetic torque under this constraint. $ \mathbf{M}_I $ and $\mathbf{M}_{I,\text{target}}$ are the atomic spin moment and the target atomic spin moment for atom $I$, respectively. The stationary point problem within the Lagrange formalism can be solved iteratively by minimizing $E_c$ with respect to $\boldsymbol\lambda_I$, $\rho(\mathbf{r})$ and $\mathbf{m}(\mathbf{r})$~\cite{2005-wu}.

This method can be extended in a straightforward manner to support a functionality that constrains only the direction of atomic spin moments. By rewriting the target atomic spin moment in Eq.~\ref{ec_deltaspin} as $\mathbf{M}_{I,\text{target}} = M_{I,\text{target}} \mathbf{e}_{I,\text{target}},$ and iteratively updating the magnitude of the target spin moment $M_{I, \text{target}} = \mathbf{M}_I(m(r))\cdot \mathbf{e}_{I,target}$ during the inner-loop calculation of optimized Lagrange multipliers $\mathbf{\lambda}_I|_{\mathbf{\lambda}_I \cdot \mathbf{e}_{I,\text{target}} = 0}$, upon the inner loop's convergence, the following conditions are satisfied: $\mathbf{M}_I\big(\mathbf{m}(\mathbf{r})\big) \parallel \mathbf{e}_{I,\text{target}},$ and $\sum_I \lambda_I\cdot M_I = 0.$ A typical example is the transition of a single atomic spin from ferromagnetic to antiferromagnetic testing, examining whether fixing the magnetic moment magnitude results in differences in energy and torque performance. The comparative studies between the full-constraint and direction-constraint method is perfomed in Fig.~S6. It is worth noting that constraining only the spin direction-rather than both magnitude and direction-reduces the number of optimization degrees of freedom, typically cutting the required convergence steps in the inner loop by approximately half. Another noteworthy aspect is that the direction-only constraint approach circumvents errors arising from discrepancies in atomic magnetic moment definitions across different software packages, thereby enabling meaningful cross-software comparisons (see Fig.~S14).

Another extension can be achieved by replacing all vector quantities in Eq.~\ref{ec_deltaspin} with scalars, enabling the method to support spin-magnitude-constrained calculations under the LSDA. It is important to clarify that the two aforementioned extensions cannot be employed simultaneously, as the spin orientation in LSDA is restricted to two discrete directions (up and down), which cannot be successfully constrained through iterative gradient-based optimization.

First, we introduce the implementation within numerical atomic orbital basis. The penalty Hamiltonian term in NAOs basis in real space can be derived from the Lagrange function as
\begin{equation}
\begin{aligned}
    H_{\mu\nu}^{\boldsymbol{\lambda}, \sigma\sigma'}(\mathbf{R}) &= \frac{\partial E_c(\rho, \{\boldsymbol{\lambda}_I\},\{\mathbf{M}_I\})}{\partial \rho^{\sigma\sigma'}_{\mu\nu}(\mathbf{R})} \\
    &= \sum_I \frac{\partial E_c(\rho, \boldsymbol{\lambda}_I,\mathbf{M}_I)}{\partial \mathbf{M}_I}\frac{\partial \mathbf{M}_I}{\partial \rho^{\sigma\sigma'}_{\mu\nu}(\mathbf{R})} \\
    &= \sum_{I}\sum_{p}  \frac{\partial ({\lambda}_I^p M_I^p)}{\partial \rho^{\sigma\sigma'}_{\mu\nu}(\mathbf{R})} \\
    &= \sum_{I}\sum_{p} {\lambda}_{I}^p\frac{\partial M_{I}^p}{\partial \rho^{p}_{\mu\nu}(\mathbf{R})}\frac{\partial \rho^{p}_{\mu\nu}(\mathbf{R})}{\partial \rho^{\sigma\sigma'}_{\mu\nu}(\mathbf{R})}.
\end{aligned}
\end{equation}
Here
\begin{equation}
\rho^p_{\mu\nu}(\mathbf{R}) = \sum_{\sigma\sigma'}\bm{\sigma}^p \rho^{\sigma\sigma'}_{\mu\nu}(\mathbf{R}),
\end{equation}
where $\mathbf{R}$ is the lattice vector between basis functions $\mu$ and $\nu$.
The last term in the chain derivative on the right-hand side is Pauli matrix, the rest of the term can continue to expand the calculation by Eq.~\ref{atomic_mag} as
\begin{equation}
    \sum_{lm}\frac{\partial M_I^{p}}{\partial n^{p}_{Ilm}}\frac{\partial n^{p}_{Ilm}}{\partial \rho^p_{\mu\nu}(\mathbf{R})}=\sum_{lm} \langle\phi^{\mathbf{0}}_{\mu} |\alpha_{Ilm}\rangle\langle\alpha_{Ilm}|\phi^{\mathbf{R}}_{\nu}\rangle.
\end{equation}

The penalty term of Hamiltonian has the form of
\begin{equation}
    H_{\mu\nu}^{\boldsymbol\lambda, \sigma\sigma'}(\mathbf{R}) = \sum_I f(I,\sigma\sigma')\sum_{lm}\langle\phi^{\mathbf{0}}_{\mu} |\alpha_{Ilm}\rangle\langle\alpha_{Ilm}|\phi^{\mathbf{R}}_{\nu}\rangle,
\end{equation}
where the 2$\times$2 parameter matrix $f(I,\sigma\sigma')$ from Pauli matrix is 
\begin{equation}
    f(I,\sigma\sigma') = 
    \left( \begin{array}{cc} \lambda_{I}^x & \lambda_{I}^x+i\lambda_{I}^y \\ \lambda_{I}^x-i\lambda_{I}^y & -\lambda_{I}^z \end{array} \right).
\end{equation}
This modified Hamiltonian ensures that the results satisfy the constraints and contribute accordingly to the total energy, and the contribution of energy from the penalty term as follows:
\begin{equation}
    E^{\bm{\lambda}} = \sum_{i}f_i\langle \psi_i|\hat{h}^{\bm{\lambda}}|\psi_i\rangle - \sum_I \bm{\lambda}_I \cdot \mathbf{M}_{I, target},
    \label{cdft-energy}
\end{equation}
where $\hat{h}^{\boldsymbol\lambda}$ is symbol of penalty operator. 

The correction terms of atomic force and lattice stress can be calculated by the derivative of the energy term in Eq.~\ref{cdft-energy}. The expression of penalty force in the direction along $p=x,y,z$ is
\begin{equation}
\begin{aligned}
    \mathbf{F}^{\boldsymbol\lambda,p}_I = - 2\sum_{\mathbf{R}}&\sum_{\mu\nu}\sum_{\sigma\sigma'}\rho_{\mu\nu}^{\sigma\sigma'}(\mathbf{R}) \times \\
    &\sum_{\mathbf{R}'}\sum_{lm}\langle\phi_\mu^\mathbf{0}|\alpha_{Ilm}^{\mathbf{R}'}\rangle f(I, \sigma\sigma')\frac{\partial}{\partial \boldsymbol\tau^p_I}\langle\alpha_{Ilm}^{\mathbf{R}'}|\phi_\nu^\mathbf{R}\rangle,
\end{aligned}
\end{equation}
where $\tau^p_I$ is the $p$-component coordinates of $I$-th atom $\boldsymbol\tau$. The penalty stress term is 
\begin{equation}
\begin{aligned}
    \Sigma^{\boldsymbol\lambda}_{\alpha\beta}=
    -\frac{1}{\Omega}&\sum_{\mathbf{R}}\sum_{\mu\nu}\sum_{\sigma\sigma'}\rho_{\mu\nu}^{\sigma\sigma'}(\mathbf{R})\sum_{I}\sum_{\mathbf{R}'}\sum_{lm}\\&\Big(\frac{\partial}{\partial \tau^{\alpha}_{\mu I}}\langle\phi_\mu^\mathbf{0}|\alpha_{Ilm}^{\mathbf{R}'}\rangle \tau^{\beta}_{\mu I} f(I, \sigma\sigma')\langle\alpha_{Ilm}^{\mathbf{R}'}|\phi_\nu^\mathbf{R}\rangle  \\
    &  +\langle\phi_\mu^\mathbf{0}|\alpha_{Ilm}^{\mathbf{R}'}\rangle f(I, \sigma\sigma')\frac{\partial}{\partial \tau^{\alpha}_{I\nu}}\langle\alpha_{Ilm}^{\mathbf{R}'}|\phi_\nu^\mathbf{R}\rangle\tau^{\beta}_{I\nu}\Big)
\end{aligned}
\end{equation}
where $\alpha, \beta$ are indices of stress $3\times3$ tensor.

For plane wave basis, directly constructing the full Hamiltonian matrix is typically infeasible due to the large number of basis functions involved. Instead, iterative diagonalization methods are employed to solve the Kohn-Sham equations, which require only the computation of the action of the Hamiltonian operator on the wave function. As a result, within the plane-wave basis set, the formula for the Hamiltonian correction term at a specified $\mathbf{k}$-point is as follows
\begin{equation}
\hat{h}^{\boldsymbol\lambda}|c^\sigma_{n\mathbf{k}}(\mathbf{G})\rangle = \sum_{I}\sum_{\sigma'} f(I,\sigma\sigma')\sum_{lm} S_{In\mathbf{k}lm}^{\sigma'}|\alpha_{Ilm}(\mathbf{k+G})\rangle.
\end{equation}
where 
\begin{equation}
S_{In\mathbf{k}lm}^{\sigma} = \langle \alpha_{Ilm}(\mathbf{k+G})|c^\sigma_{n\mathbf{k}}(\mathbf{G})\rangle
\end{equation}
is the overlap of projected orbital and wave function.

The terms for the correction of atomic force and lattice stress can be also derived from the energy term of the Eq.~\ref{cdft-energy} through its differentiation. The expression for atomic forces is
\begin{equation}
\begin{aligned}
\bm{F}^{\boldsymbol\lambda,p}_I = -2\sum_{\mathbf{k}}&\sum_{I'}\sum_{lm}\sum_{\sigma\sigma'}f(I', \sigma\sigma')S_{I'n\mathbf{k}lm}^{\sigma} \\
&\langle \alpha_{I'lm}(\mathbf{G+k})|(-j)(\mathbf{G+k})_p |c^{\sigma'}_{n\mathbf{k}}(\mathbf{G})\rangle,    
\end{aligned}
\end{equation}
and for stress is
\begin{equation}
\Sigma^{\boldsymbol\lambda}_{\alpha\beta} = -\frac{2}{\Omega}\sum_{\mathbf{k}}\sum_{I'}\sum_{lm}\sum_{\sigma\sigma'}f(I', \sigma\sigma')S_{I'n\mathbf{k}lm}^{\sigma} \frac{\partial S_{I'n\mathbf{k}lm}^{\sigma'}}{\partial \varepsilon_{\alpha\beta}},
\end{equation}
where $\Omega$ is the cell volume, $\varepsilon_{\alpha\beta}$ is the lattice strain tensor, and detail of last derivative term is similar with nonlocal pseudopotential term in Ref.~\cite{Nelson1985}.

\begin{figure}
    \centering
    \includegraphics[width=0.7\linewidth]{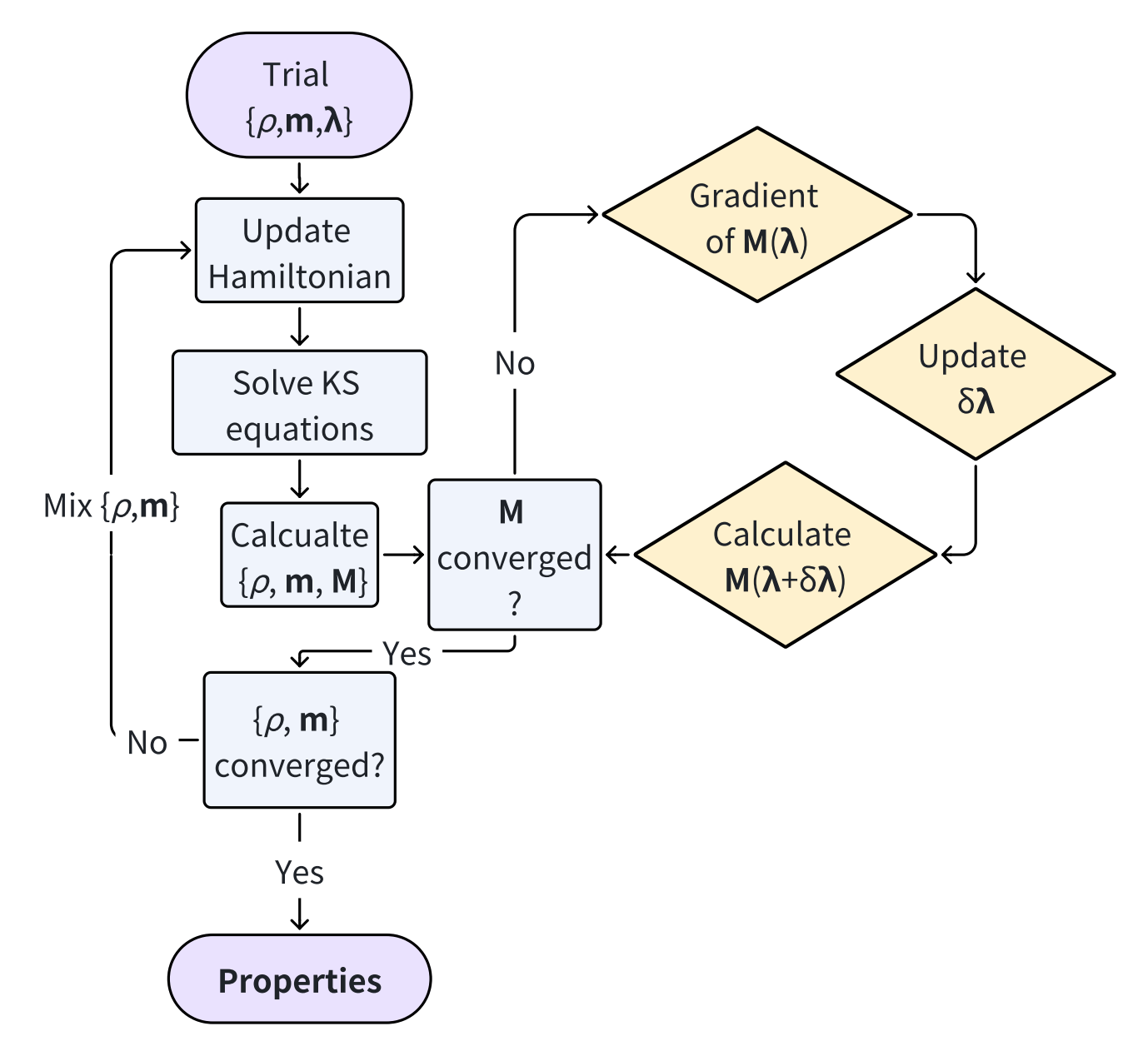}
    \caption{\textbf{The double-loop workflow of spin-cDFT.} Framework of variant Lagrange quantities spin-cDFT algorithm implemented in ABACUS for both PW and NAOs basis. The electron density is $\rho$ and the spin density vector is $\mathbf{m}$. The atomic magnetization is $\mathbf{M}$.}
    \label{fig:dspin_framework}
\end{figure}

We present a uniform flowchart of the specific implementation of spin-cDFT in Fig.~\ref{fig:dspin_framework} for both PW and NAO basis. As depicted in Fig.~\ref{fig:dspin_framework}, this procedure necessitates an additional iteration within the self-consistent field (SCF) cycle of DFT to ascertain the fixed-point solution satisfying the spin constraints. The process involves solving the Kohn-Sham equations with a given set of Lagrange multipliers, which are then updated to minimize the deviation between the calculated and target magnetic moments. This minimization is typically achieved through a conjugate-gradient scheme or a similar optimization technique, ensuring that the magnetic moments converge to the targets within a predefined tolerance. The iterative loop continues, alternating between the solution of the Kohn-Sham equations and the optimization of the Lagrange multiplier, until the system reaches a self-consistent solution where the constraints are satisfied to a desired level of accuracy. Through comparative testing, we found that the energy calculation results from the adaptive algorithm implemented in this work demonstrate agreement with the penalty function methods employed in other software packages (see Fig.~S14). 
\revise{Compared to the penalty function method, which exhibits nonlinear errors dependent on $\lambda$ across different excited states (Fig.~S14), new method demonstrates consistent accuracy and is suitable for large-scale sampling.}

We observed that in calculations of the Fe system, the optimization of lambda typically introduces an inner-loop iteration count on the order of 5. Since the inner loop also requires solving eigenvalue problems, this leads to an approximately fivefold increase in total computational time when using numerical atomic orbital basis sets \revise{(Fig.~S15)}. For plane-wave basis set, due to its high dimensionality, the inner loop employs a single-step subspace solution of the Davidson diagonalization algorithm when the magnetic moment error is large, rather than the full Davidson solver used in the outer loop. This merit significantly improves the efficiency of magnetic moments in the inner loop while guaranteeing the correctness of final magnetic moment convergence. In tests, each step of the inner loop with PW basis sets consumes significantly less time than the outer loop, resulting in only a modest increase ($\sim 1/3$) in total computational time (see Fig.~S7).
\revise{Tests on systems of varying sizes (see Fig.~S15) demonstrate that for both LCAO and PW basis implementations, the proportion of additional computational cost introduced by cDFT remains nearly constant as the system size increases. This indicates that the current cDFT method does not alter the scaling of the original SCF calculation; the extra cost only increases the prefactor of the total computational time, ensuring its scalability. Furthermore, the present method, due to its strict enforcement of magnetic moment constraints at each electronic step, reduces the degrees of freedom of the electronic state during the SCF process, leading to superior convergence speed compared to the penalty method (see Table.~S6).}

\subsection{First-principles Calculations}
All calculations in this work were performed with ABACUS, utilizing plane waves or numerical atomic orbitals~\cite{Li2016} to expand wavefunctions. The pseudopotentials were sourced from the Pseudo-Dojo pseudopotential library~\cite{PseudoDojo}. The numerical atomic orbitals of ABACUS are generated by optimizing the spillage function averaged over structures; this strategy was originally proposed in the works of Chen {\it et al.}~\cite{Chen2010} (denoted as v1.0) and improved by Lin {\it et al.}~\cite{lin_accuracy_2020} by including the kinetic terms (denoted as v2.0). In this work, a numerically further improved version is also used (denoted as v2.1).

These v2.1 orbitals have not yet been officially released on the official website but the generation code is online available~\cite{orb-v2.1}. This paper performs NAOs calculations using v2.0 and v2.1 orbitals, including Single-$\zeta$ (SZ), Double-$\zeta$ plus polarization functions (DZP), and Triple-$\zeta$ plus double polarization functions (TZDP) with different cutoffs. The convergence test for the plane-wave cutoff energy is presented in the Supplementary Information. By adopting a criterion of energy error per atom below 1 meV and stress error below 0.1 kbar, the cutoff energy was set to 100 Ry for all calculations (see Fig.~S4). The Brillouin zone was uniformly sampled at 0.14 Bohr$^{-1}$ intervals. Specifically for the BCC-Fe supercell with 16 iron atoms, this k-spacing corresponds to a $5\times5\times5$ Monkhorst-Pack grid. The projected orbitals used to calculate magnetic moments were modified numerical atomic orbitals, with a modulation radius of 3 Bohr. Each inner loop of $\bm{\lambda}$ optimization converges when the maximum atomic magnetism variation is less than $\delta\mathbf{M}<10^{-7} \mu_B$, while the SCF convergence until density difference $\delta\rho<10^{-6}$. All noncollinear calculations incorporated the spin-orbit coupling (SOC) effect. We would like to clarify that while SOC effects in pure Fe are relatively weak—smaller than the intrinsic errors of our model—we deliberately retained SOC calculations to rigorously test the stability of our implementation under combined noncollinear magnetism, SOC, and cDFT conditions, as well as to validate the robustness of the entire model workflow.
\revise{The cDFT method implemented in ABACUS, capable of precisely obtaining first-principles results for arbitrary magnetic moment configurations, also serves as an effective tool for studying the systems where SOC plays a decisive role in determining the magnetic ground state or emergent phenomena~\cite{2023-yang-first}—such as heavy-element magnets, topological spin textures, or Dzyaloshinskii–Moriya interaction (DMI) driven systems. In Table S4 of the Supplementary Material, we have included a short example using the four-states method~\cite{2011-xiang-predicting} to calculate the DMI effect in the Fe–Pt system, demonstrating this capability.}

\subsection{Molecular Dynamics Simulation}
The molecular dynamics simulations utilize a modified version of LAMMPS using the TSPIN method involved simultaneous integration of lattice and spin degrees of freedom within a unified Nosé–Hoover chain (NHC) thermostat framework~\cite{2025-huang-prep}. The dynamics are propagated using symplectic integration schemes derived from an extended Hamiltonian formulation, which preserves symplecticity and ensures accurate sampling of canonical (NVT) and isothermal–isobaric (NPT) ensembles. 

The Lagrangian that governs the system’s evolution can be expressed as:
\begin{equation}
L = \sum_i\frac{p_{si}^2}{2\mu_i} + \sum_i\frac{p_i^2}{2m_i} -U(\{\mathbf{R}\}, \{\mathbf{M}\}),
\end{equation}
where $p_{si}$ and $p_{i}$ are the generalized momenta associated with the spins and the lattice positions, respectively. $\mu_i$ denotes the virtual magnetic mass. The energy function $U(\{\mathbf{R}\}, \{\mathbf{M}\})$ accounts for the interactions between the lattice, spins, and spin-lattice couplings. The Euler-Lagrange equations derived from this Lagrangian provide the following equations of motion: 
\begin{equation}
\begin{aligned}
\dot{\mathbf{R}}_i &= \frac{p_i}{m_i}, \quad &\dot{p}_i = -\frac{\partial U}{\partial \mathbf{R}_i}=\mathbf{F}_i, \\
\dot{\mathbf{M}}_i &= \frac{p_{si}}{\mu_i}, \quad &\dot{p}_{si} = -\frac{\partial U}{\partial \mathbf{M}_i} = \boldsymbol{\lambda}_i.
\end{aligned}
\end{equation}
The above equations describe the evolution of the lattice ($\mathbf{R}_i$) and spin ($\mathbf{M}_i$), where the lattice evolves according to the atomic forces ($\mathbf{F}_i$), while the spins evolve under the magnetic torque ($\boldsymbol{\lambda}_i$). For the NVT and NPT ensembles, we use the equations of motion with thermostat variables for temperature control shown in Ref.~\cite{2025-huang-prep}.

\subsection{Finite Difference Tests}
For the force test, we randomly selected one iron atom from a BCC-Fe$_{16}$ configuration and perturbed its z-coordinate with a step size of $\Delta=0.01~\text{Bohr}$, calculating the energies of 11 perturbed configurations $E(z_0+i\Delta)$. Using the finite difference formula:
\begin{equation}
   \mathbf{F}_z(i) = \frac{\partial E}{\partial \mathbf{r}_z} \approx \frac{f((i+1)\Delta)-f(-(i-1)\Delta)}{2\Delta}, 
\end{equation}
we computed numerical solutions for 9 points (excluding the edge points).

For the magnetic torque test, we randomly selected a FePt configuration and perturbed the z-component magnetic moment of one Fe atom with a step size of $\Delta=0.1~\mu_B$. We calculated the total energies of 11 perturbed configurations and determined the numerical solutions $\frac{\partial E}{\partial \mathbf{m}_z}$ using the central difference method.

We randomly selected a NiMnTi configuration to conduct stress finite-difference testing. The stress can be obtained through strain perturbation:
\begin{equation}
\sigma_{\alpha\beta} = -\frac{1}{\Omega}\frac{\partial E}{\partial \varepsilon_{\alpha\beta}}, 
\end{equation}
where $\Omega$ represents the volume, and the stress tensor $\sigma$ has six independent components: $\sigma_{11}$, $\sigma_{12}$, $\sigma_{13}$, $\sigma_{22}$, $\sigma_{23}$, $\sigma_{33}$. For each component, we constructed corresponding strained configurations. Taking $\sigma_{11}$ as an example, we generated 11 configurations with $\varepsilon_{11}=i*\Delta$ (step size $\Delta=0.0001$). According to the following formula, the strain induces the corresponding unit cell modifications:
\begin{equation}
\begin{aligned}
&\varepsilon_{11}:\begin{pmatrix}a_{11}+\varepsilon_{11} a_{11}&a_{12}&a_{13}\\a_{21}+\varepsilon_{11} a_{21}&a_{22}&a_{23}\\a_{31}+\varepsilon_{11} a_{31}&a_{32}&a_{33}\end{pmatrix}
&&\varepsilon_{12}:\begin{pmatrix}a_{11}+\varepsilon_{12} a_{12}&a_{12}&a_{13}\\a_{21}+\varepsilon_{12} a_{22}&a_{22}&a_{23}\\a_{31}+\varepsilon_{12} a_{32}&a_{32}&a_{33}\end{pmatrix}\\
&\varepsilon_{22}:\begin{pmatrix}a_{11} &a_{12}+\varepsilon_{22} a_{12}&a_{13}\\a_{21} &a_{22}+\varepsilon_{22} a_{22}&a_{23}\\a_{31} &a_{32}+\varepsilon_{22} a_{32}&a_{33}\end{pmatrix}
&&\varepsilon_{13}:\begin{pmatrix}a_{11}+\varepsilon_{13} a_{13}&a_{12}&a_{13}\\a_{21}+\varepsilon_{13} a_{23}&a_{22}&a_{23}\\a_{31}+\varepsilon_{13} a_{33}&a_{32}&a_{33}\end{pmatrix}\\
&\varepsilon_{33}:\begin{pmatrix}a_{11} &a_{12}&a_{13}+\varepsilon_{33} a_{13}\\a_{21} &a_{22}&a_{23}+\varepsilon_{33} a_{23}\\a_{31} &a_{32}&a_{33}+\varepsilon_{33} a_{33}\end{pmatrix}
&&\varepsilon_{23}:\begin{pmatrix}a_{11} &a_{12}+\varepsilon_{23} a_{13}&a_{13}\\a_{21} &a_{22}+\varepsilon_{23} a_{23}&a_{23}\\a_{31} &a_{32}+\varepsilon_{23} a_{33}&a_{33}\end{pmatrix}
\end{aligned}
\end{equation}
The raw data is shown in the Table.~S1/S2/S3 of Supplementary Information.

\backmatter

\section*{Data availability}
\revise{The data and models used in this study have been made open-source and are hosted on AIS Square (https://www.aissquare.com/), where they can be accessed online (https://www.aissquare.com/datasets/detail?pageType=datasets\&name=Fe-DeepSpin\&id=386).}

\section*{Code availability}
The codes used to produce the results are available from the corresponding author upon reasonable request. The relevant implementation is scheduled to be open-sourced in the next major release of ABACUS and may be accessed at: https://github.com/deepmodeling/abacus-develop.

\section*{Author contributions}
D.Z. and W.Z. designed and implemented the spin-constrained method in ABACUS. Z.C. participated in the discussion, design, and benchmark of the implementation. Y.H. and L.Z. provided guidance on the projection method. X.P. and Y.W. constructed the first-principles database and trained the iron model. D.Z. and Z.H. implemented code for magnetic model training and molecular dynamics simulations. W.Z. and D.Z. wrote the main manuscript. W.Z., B.X. and M.C. conceived and directed the research, supervised the analysis and interpretation. All authors contributed to the discussions and the final editing.

\section*{Competing interests}
The authors declare no competing financial or non-financial interests.

\section*{Acknowledgements}
We thank Zuxin Jin and Han Wang for many helpful discussions. We gratefully acknowledges AI for Science Institute, Beijing (AISI). W.Z. gratefully acknowledges support from the Hongyi postdoctoral fellowship of Wuhan University. M.C. gratefully acknowledges funding support from the National Natural Science Foundation of China (grant no. 12122401,12074007,12135002).

\hypersetup{
    hidelinks
}
\bibliography{ref}

\end{document}


\title[Article Title]{Supplementary information: Integrating Deep-Learning-Based Magnetic Model and Non-Collinear Spin-Constrained Method: Methodology, Implementation and Application}

\author[1]{\fnm{Daye} \sur{Zheng}}

\author[1]{\fnm{Xingliang} \sur{Peng}}

\author[1]{\fnm{Yike} \sur{Huang}}

\author[1]{\fnm{Yinan} \sur{Wang}}

\author[1]{\fnm{Duo} \sur{Zhang}}

\author[2,3]{\fnm{Zhengtao} \sur{Huang}}

\author[8]{\fnm{Zefeng} \sur{Cai}}

\author[4,1]{\fnm{Linfeng} \sur{Zhang}}

\author*[1,5]{\fnm{Mohan} \sur{Chen}}\email{mohanchen@pku.edu.cn}

\author*[6]{\fnm{Ben} \sur{Xu}}\email{bxu@gscaep.ac.cn}

\author*[1,7]{\fnm{Weiqing} \sur{Zhou}}\email{weiqingzhou@whu.edu.cn}

\affil[1]{AI for Science Institute, Beijing 100080, P. R. China}
\affil[2]{Graduate School of China Academy of Engineering Physics, Beijing 100088, P. R. China}
\affil[3]{International School of Materials Science and Engineering, Wuhan University of Technology, Wuhan 430070, P. R. China}
\affil[4]{DP Technology, Beijing 100080, P. R. China}
\affil[5]{HEDPS, CAPT, School of Physics and College of Engineering, Peking University, Beijing, 100871, P. R. China}
\affil[6]{Graduate School of China Academy of Engineering Physics, Beijing 100193, P. R. China}
\affil[7]{Key Laboratory of Artificial Micro- and Nano-structures of Ministry of Education and School of Physics and Technology, Wuhan University, Wuhan 430072, P. R. China}
\affil[8]{Department of Materials Science and Engineering, Carnegie Mellon University, Pittsburgh, PA 15213, USA}

\maketitle

\newpage

\begin{figure*}
    \centering    
    \includegraphics[width=1.0\textwidth]{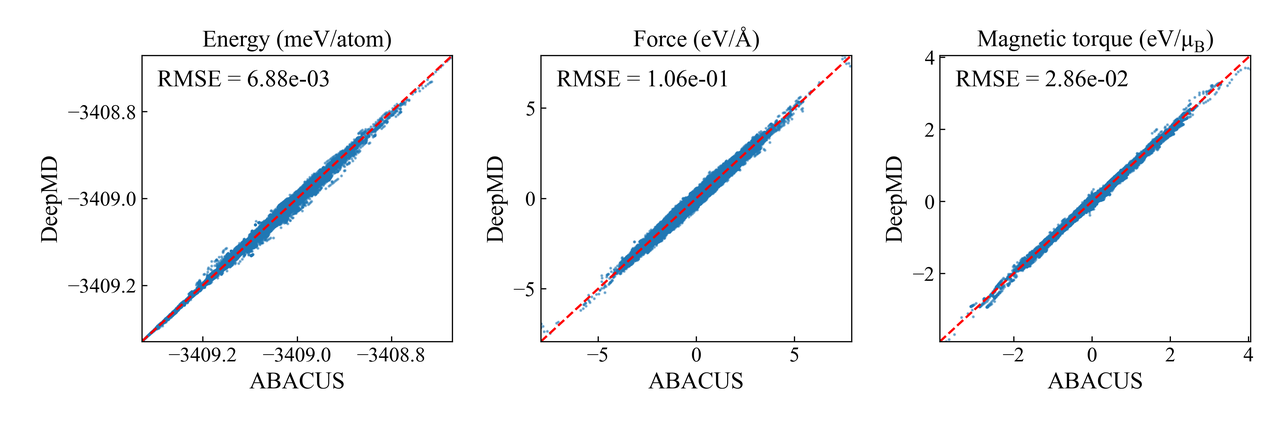}
    \caption{
    \textbf{The training error.}
    For DZP model, the training error of energy, atomic force, and magnetic torque. RMSE denotes Root Mean Square Error.}
    \label{fig:}
\end{figure*}

\begin{figure*}
    \centering    
    \includegraphics[width=1.0\textwidth]{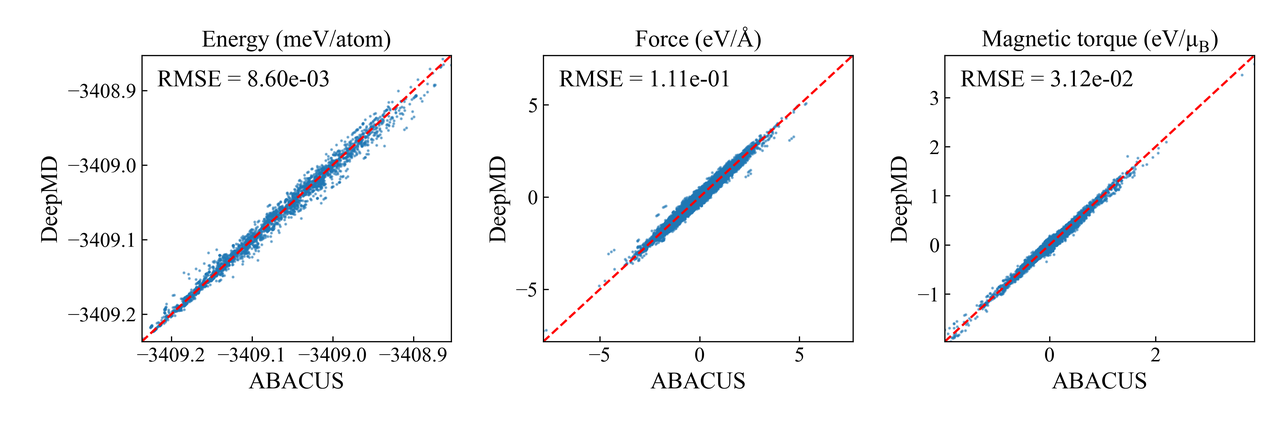}
    \caption{
    \textbf{The prediction error.}
    For DZP model, the prediction error of energy, atomic force, and magnetic torque for the test set. RMSE denotes Root Mean Square Error.}
    \label{fig:}
\end{figure*}

\begin{figure*}
    \centering    
    \includegraphics[width=1.0\textwidth]{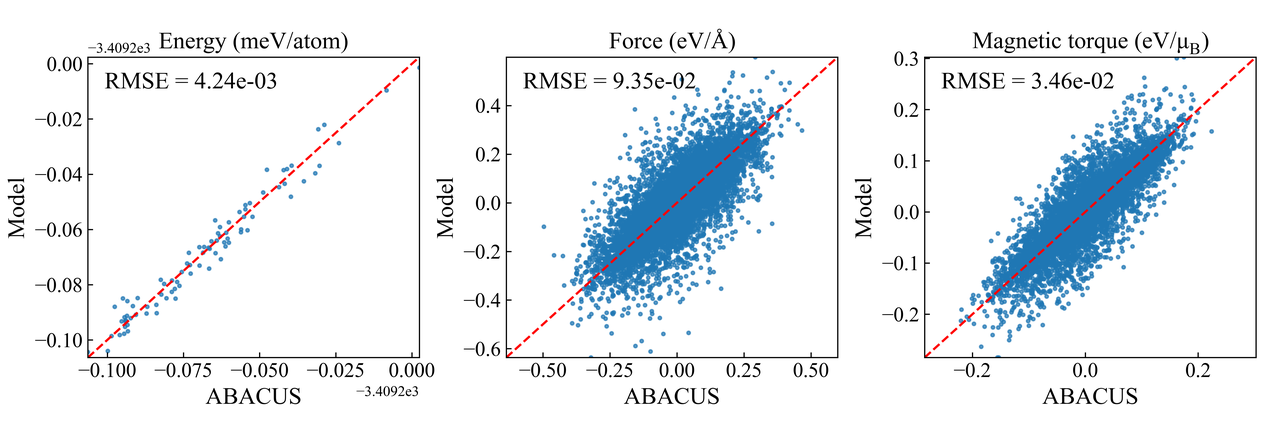}
    \caption{
    \textbf{Scalability of the model.}
    The prediction error of the DZP model for the systems with 32 atoms. We randomly generated 100 BCC structures each containing 32 atoms, with atomic magnetic moment directions randomly rotated up to \ang{90}, atomic positions randomly displaced within 0.02 $\rm \AA$, and atomic magnetic moment magnitudes randomly perturbed by up to 0.2 $\mu_B$. The model was then used to predict the energy, atomic forces, and magnetic torque for these structures, and the results were systematically compared with DFT calculations.}
    \label{fig:}
\end{figure*}

\begin{table}[htbp]
\centering
\caption{\textbf{Atomic Force Finite difference test for BCC-Fe.} We randomly selected one iron atom from a BCC-Fe$_{16}$ configuration and perturbed its z-coordinate with a step size of $\Delta=0.01~\text{Bohr}$.}
\label{tab:fd_test_force}
\begin{tabular}{
    S[table-format=-1.4] 
    S[table-format=-7.14] 
    S[table-format=-1.6] 
    S[table-format=-1.6] 
    S[table-format=1.6] 
}
\toprule
{$i\Delta$ (\si{\angstrom})} & 
{$E_i$ (\si{eV})} & 
{$F_{\text{analytic}}$ (\si{eV/\angstrom})} & 
{$F_{\mathrm{FD}}$ (\si{eV/\angstrom})} & 
{$|F_{\text{analytic}}-F_{\mathrm{FD}}|$ (\si{eV/\angstrom})} \\
\midrule
-0.0211 & -54548.69361010315   &  0.185463 &  0.183174 & 0.002289 \\
-0.0158 & -54548.694319595874  &  0.104196 &  0.098514 & 0.005682 \\
-0.0105 & -54548.6946527533    &  0.022884 &  0.026653 & 0.003770 \\
-0.0052 & -54548.69460168539   & -0.057813 & -0.052678 & 0.005140 \\
 0.0000 & -54548.69409522063   & -0.138331 & -0.139361 & 0.001030 \\
 0.0052 & -54548.6931267173    & -0.218926 & -0.221314 & 0.002388 \\
 0.0105 & -54548.69175288521   & -0.299433 & -0.297749 & 0.001680 \\
 0.0158 & -54548.68997541114   & -0.379551 & -0.379952 & 0.000401 \\
 0.0211 & -54548.68773156097   & -0.459275 & -0.454248 & 0.005030 \\
\bottomrule
\end{tabular}
\end{table}

\begin{table}[htbp]
\centering
\caption{\textbf{Magnetic Torque Finite difference test for FePt.} We randomly selected a FePt configuration and perturbed the z-component magnetic moment of one Fe atom with a step size of $\Delta=0.1~\mu_B$.}
\label{tab:fd_test_magnetic}
\begin{tabular}{
    S[table-format=-1.1] 
    S[table-format=-7.14] 
    S[table-format=-1.6] 
    S[table-format=-1.6] 
    S[table-format=1.6] 
}
\toprule
{$i\Delta$ ($\mu_{\mathrm{B}}$)} & 
{$E_i$ (\si{eV})} & 
{$\lambda_{\text{analytic}}$ (\si{eV/\mu_{\mathrm{B}}})} & 
{$\lambda_{\mathrm{FD}}$ (\si{eV/\mu_{\mathrm{B}}})} & 
{$|\lambda_{\text{analytic}}-\lambda_{\mathrm{FD}}|$ (\si{eV/\mu_{\mathrm{B}}})} \\
\midrule
-0.4 & -13934.14190847913 &  0.254131 &  0.252482 & 0.001649 \\
-0.3 & -13934.1654258067  &  0.214478 &  0.212477 & 0.002001 \\
-0.2 & -13934.18440389591 &  0.162852 &  0.160217 & 0.002635 \\
-0.1 & -13934.19746931246 &  0.095753 &  0.092625 & 0.003128 \\
 0.0 & -13934.20292906196 &  0.010113 &  0.006368 & 0.003745 \\
 0.1 & -13934.19874304215 & -0.097961 & -0.102389 & 0.004428 \\
 0.2 & -13934.18245109053 & -0.232631 & -0.237618 & 0.004987 \\
 0.3 & -13934.15121933604 & -0.396904 & -0.402063 & 0.005159 \\
 0.4 & -13934.10203837163 & -0.592095 & -0.597729 & 0.005634 \\
\bottomrule
\end{tabular}
\end{table}

\begin{table}[htbp]
\centering
\caption{\textbf{Cell Stress Finite difference test for NiMnTi.} For each component, we constructed corresponding strained configurations. Taking $\sigma_{11}$ as an example, we generated 11 configurations with $\varepsilon_{11}=i*\Delta$ (step size $\Delta=0.0001$).}
\label{tab:strain_diff}
\begin{tabular*}{\linewidth}{
    @{\extracolsep{\fill}}
    S[table-format=-1.4]
    S[table-format=1.2]
    S[table-format=1.2]
    S[table-format=1.2]
    S[table-format=1.2]
    S[table-format=1.2]
    S[table-format=1.2]
    @{}
}
\toprule
\multicolumn{1}{c}{$i\Delta$} & 
\multicolumn{6}{c}{$\text{abs}(\varepsilon_{\text{analytic}}-\varepsilon_{\text{FD}})$ (kpar)} \\
\cmidrule(lr){2-7}
& {$\varepsilon_{11}$} & {$\varepsilon_{12}$} & {$\varepsilon_{21}$} & {$\varepsilon_{22}$} & {$\varepsilon_{23}$} & {$\varepsilon_{33}$} \\
\midrule
-0.0004 & 0.15 & 0.03 & 0.06 & 0.06 & 0.03 & 0.22 \\
-0.0003 & 0.13 & 0.07 & 0.06 & 0.07 & 0.02 & 0.23 \\
-0.0002 & 0.11 & 0.06 & 0.05 & 0.06 & 0.03 & 0.22 \\
-0.0001 & 0.08 & 0.06 & 0.02 & 0.04 & 0.05 & 0.20 \\
 0.0000 & 0.05 & 0.07 & 0.05 & 0.05 & 0.03 & 0.18 \\
 0.0001 & 0.04 & 0.08 & 0.02 & 0.07 & 0.00 & 0.20 \\
 0.0002 & 0.00 & 0.09 & 0.01 & 0.07 & 0.04 & 0.18 \\
 0.0003 & 0.00 & 0.12 & 0.00 & 0.09 & 0.05 & 0.16 \\
 0.0004 & 0.04 & 0.10 & 0.00 & 0.09 & 0.01 & 0.16 \\
\bottomrule
\end{tabular*}
\end{table}

\begin{figure*}
    \centering    
    \includegraphics[width=1.0\textwidth]{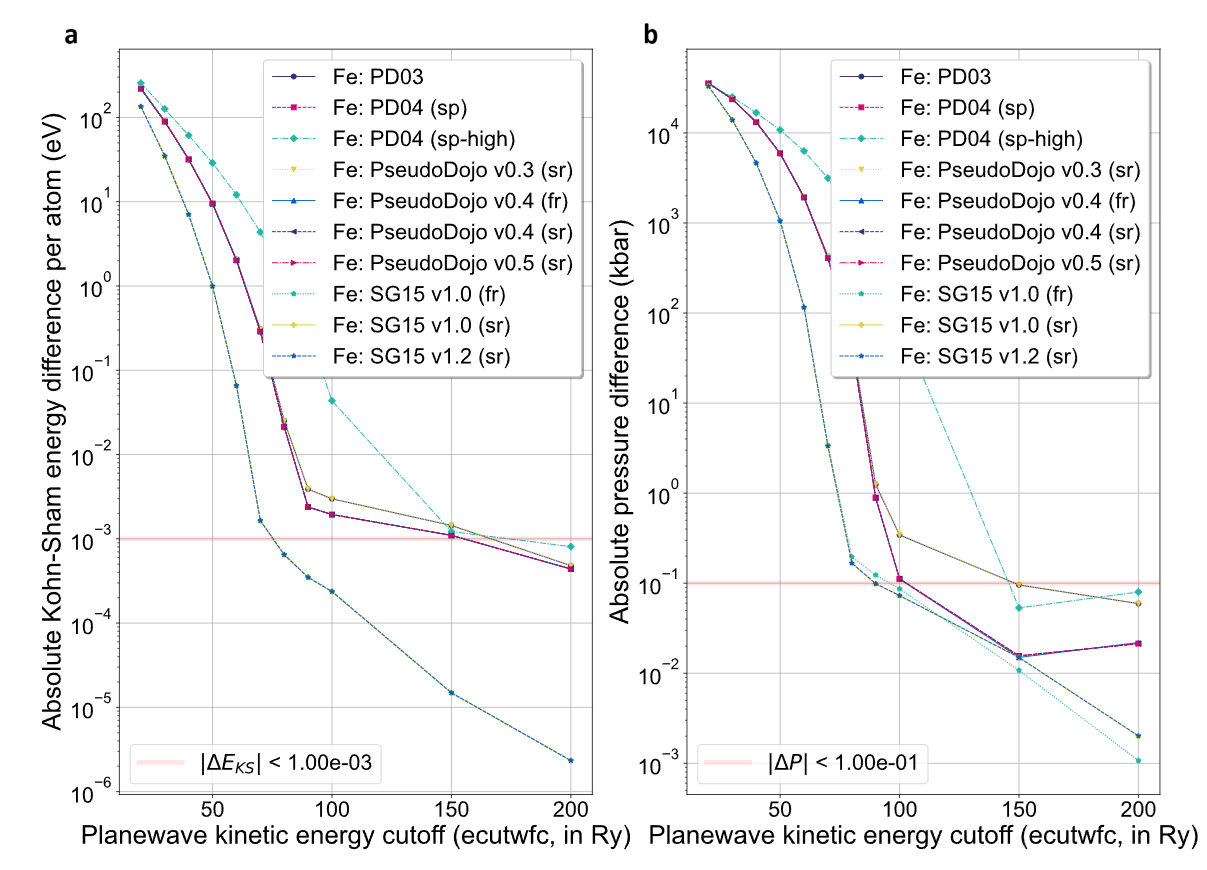}
    \caption{\textbf{Convergence tests of plane-wave cutoff energy for pseudopotentials.} The energy errors (a) and stress errors (b) versus cutoff energy for different optional norm-conserving pseudopotentials of Fe. The DoJo-fr pseudopotential is used in this work.}
    \label{fig:}
\end{figure*}

\begin{figure*}
    \centering    
    \includegraphics[width=1.0\textwidth]{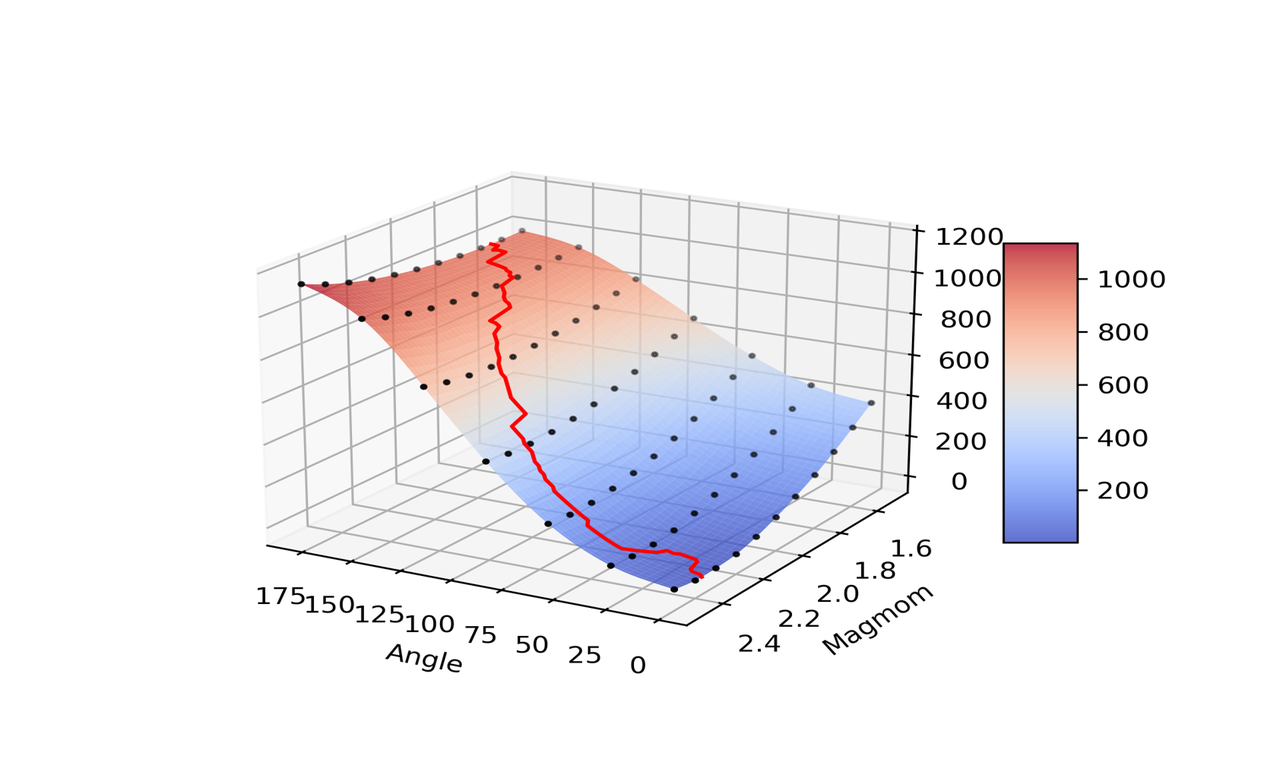}
    \caption{\textbf{Two-dimensional magnetic potential energy surface.}
    The magnetic potential energy surface of BCC-Fe includes both transverse and longitudinal perturbations, where $\theta$ is the angle between the magnetic moment of two neighbor Fe atom, magmon is the $|\mathbf{M}|$ in the unit of $\mu_B$. The colorbar denotes the relative energy with FM configuration in the unit of meV. }
    \label{fig:}
\end{figure*}

\begin{figure*}
    \centering    
    \includegraphics[width=0.8\textwidth]{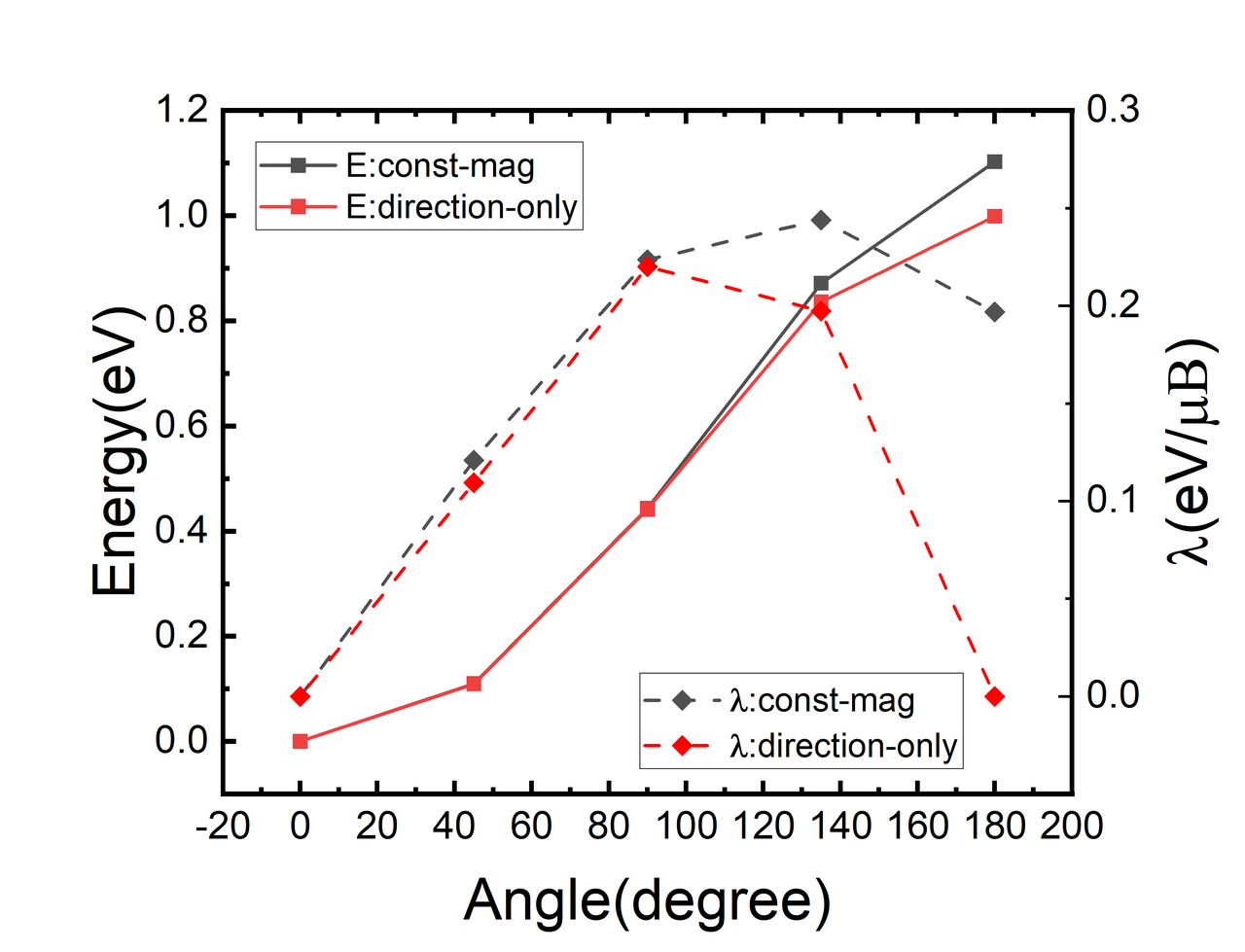}
    \caption{
    \textbf{Comparative studies between the full-constraint and direction-constraint methods in BCC-Fe.} 
    In the full-constraint approach, we fixed the magnetic moment magnitude $|\mathbf{M}|$ at the FM ground state value while varying the angle between neighboring moments to 0°, 45°, 90°, 135°, and 180°. Parallel calculations were carried out using the direction-constraint method for identical configurations. The black curves represent results from the full-constraint method, while the red curves correspond to the direction-constraint method, with solid lines denoting relative energy and dashed lines indicating magnetic torque. The results demonstrate good agreement between both methods for angles below 90°, confirming the stability of the FM ground state in this regime. However, for angles exceeding 90°, significant deviations emerge as the direction-constraint method allows the moment magnitude to relax toward the antiferromagnetic ground state, which exhibits a smaller moment than the FM state. }
    \label{fig:}
\end{figure*}

\begin{figure*}
    \centering    
    \includegraphics[width=1.0\textwidth]{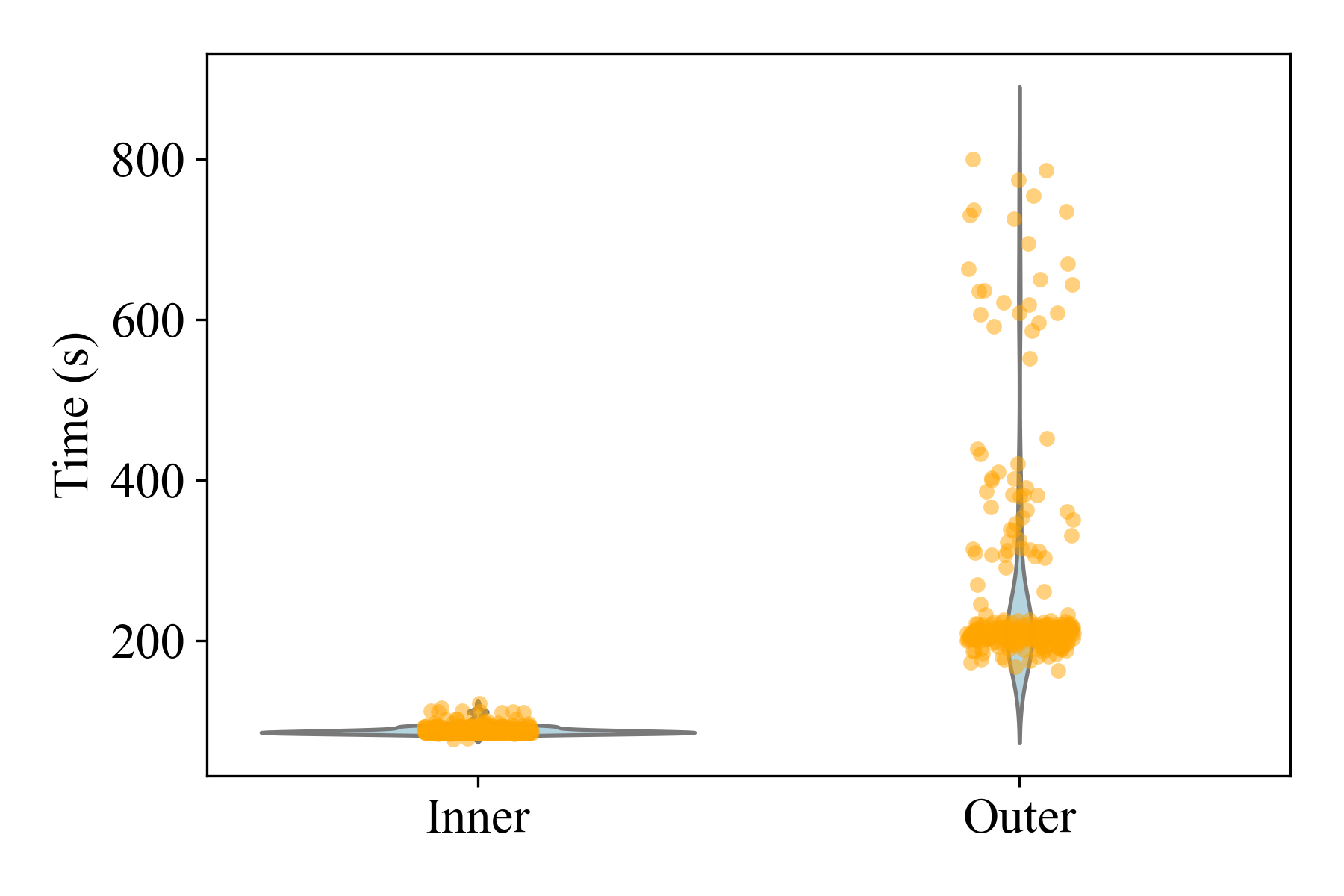}
    \caption{
    \textbf{The efficiency analysis.}
    The analysis of the time consumption of the inner loop (Inner) and the regular SCF calculation (Outer, obtained by subtracting the inner loop time from the total SCF step time) for 8 FCC Fe16 configurations using PW basis. The horizontal distribution here does not represent actual physical meaning—the width of the colored blocks corresponds to the counting density of time. The average time for the inner loop was 88s, while the regular SCF calculation took 267s. This means the computational cost of Deltaspin increases by approximately 1/3 compared to a standard SCF calculation.}
    \label{fig:}
\end{figure*}

\begin{figure*}
    \centering    
    \includegraphics[width=1.0\textwidth]{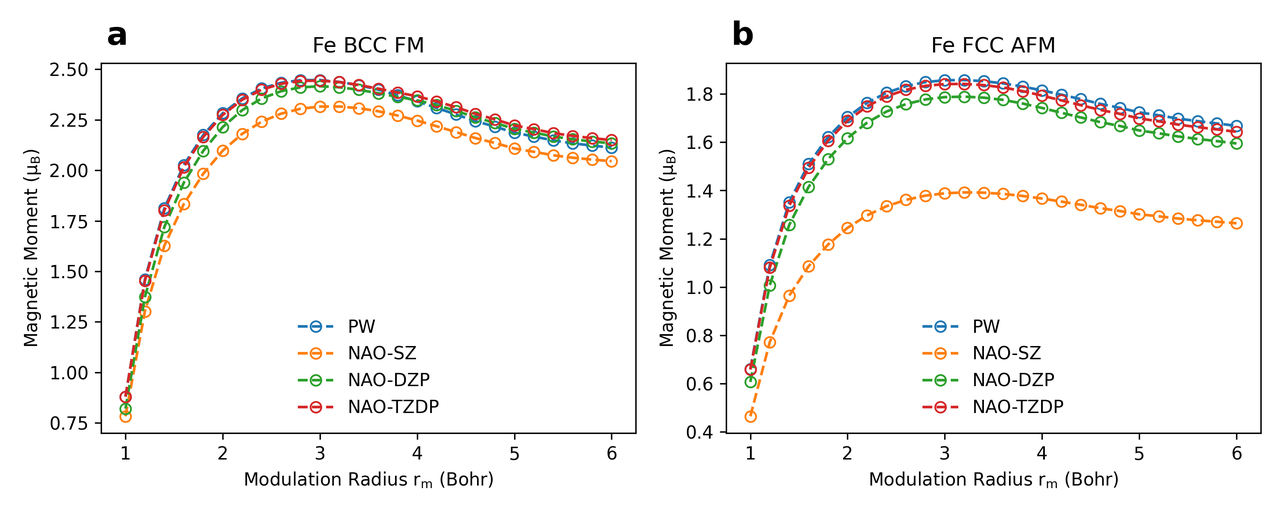}
    \caption{
    \textbf{The basis sensitivity for modulation radius.}
    The projected results as a function of modulation radius for different PW/NAO basis sets (using SZ/DZP/TZDP orbitals with 8 au cutoff, respectively). PW calculations employed DZP orbitals with 8 au cutoff for projection. (a) BCC Fe FM phase, (b) FCC Fe AFM phase.}
    \label{fig:}
\end{figure*}

\begin{figure*}
    \centering    
    \includegraphics[width=1.0\textwidth]{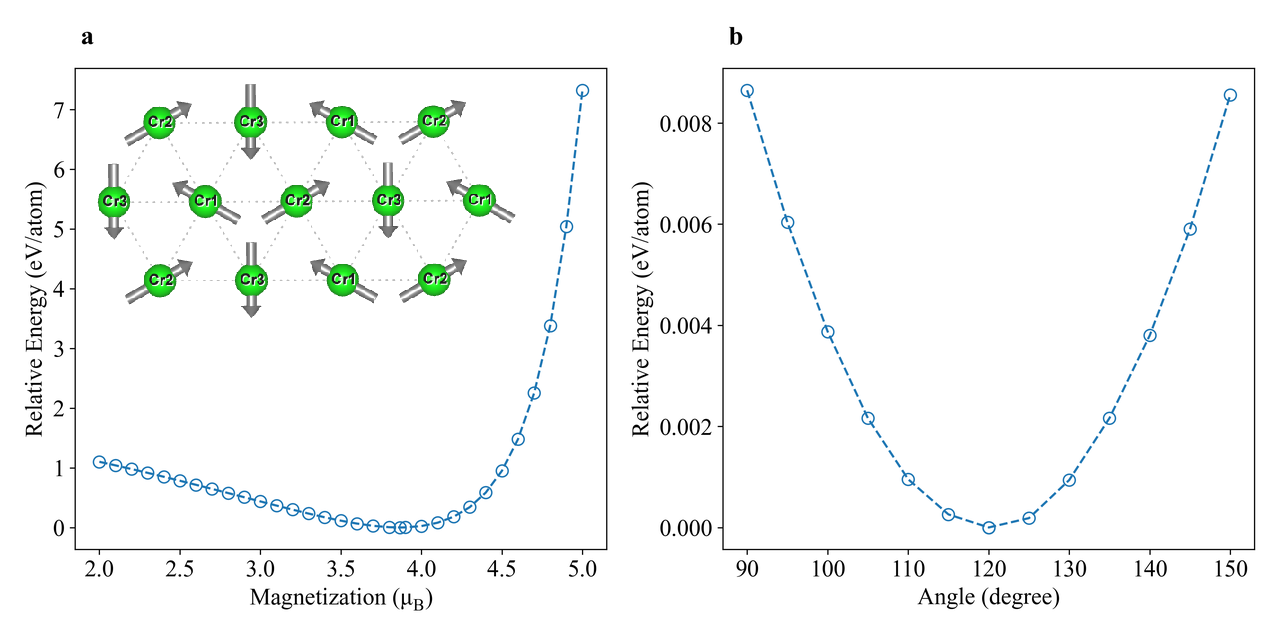}
    \caption{
    \textbf{Deltaspin calculation tests for Cr$_3$ non-collinear system.} 
    The ground state of Cr$_3$ system is obtained by a unconstrained non-collinear calculations using DZP basis, obtaining a ground-state magnetic moment of 3.866 $\mu_B$ with \ang{120} angles between the moments of the three Cr atoms. Then the constraint is introduced. a) Fixed spin orientation, varying moment magnitude: The energy minimum occurs at 3.866 $\mu_B$, consistent with the ground-state calculation. b) Fixed moment magnitude, rotating spin orientation: With all moments constrained to a common plane, the energy minimum appears when the angle between the first and second Cr moments is \ang{120} (horizontal axis shows this angular variation).}
    \label{fig:}
\end{figure*}

\begin{figure*}
    \centering    
    \includegraphics[width=1.0\textwidth]{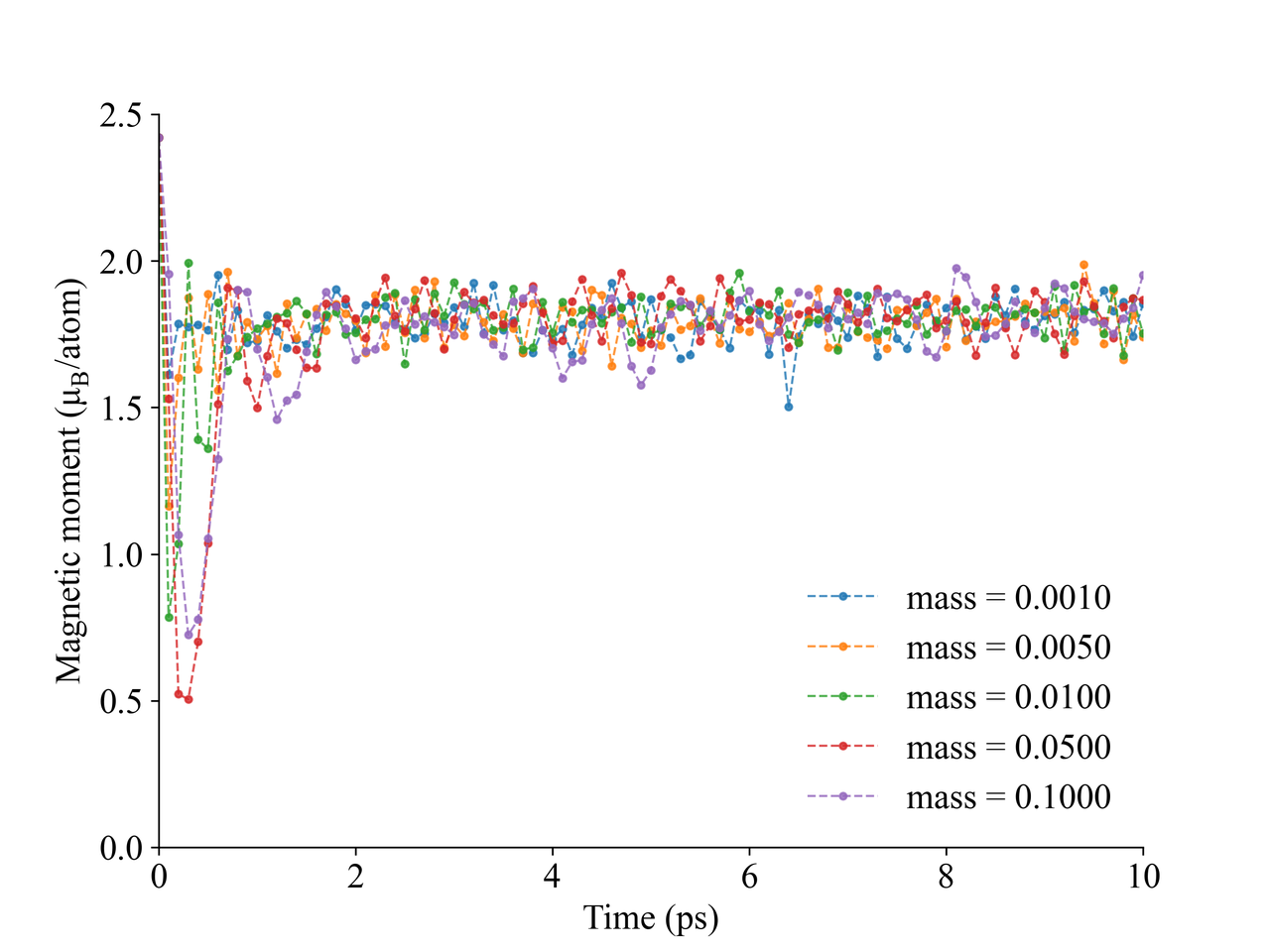}
    \caption{
    \textbf{The effect of virtual magnetic mass.}
    The evolution of average magnetic moment ($|\sum_{i}^{N_{atom}}{m_{i}}|/N_{atom}$) of BCC-Fe for various virtual magnetic mass. The MD simulations are performed on BCC-Fe using an NVT system at a temperature of 700K.}
    \label{fig:}
\end{figure*}

\begin{figure*}
    \centering    
    \includegraphics[width=1.0\textwidth]{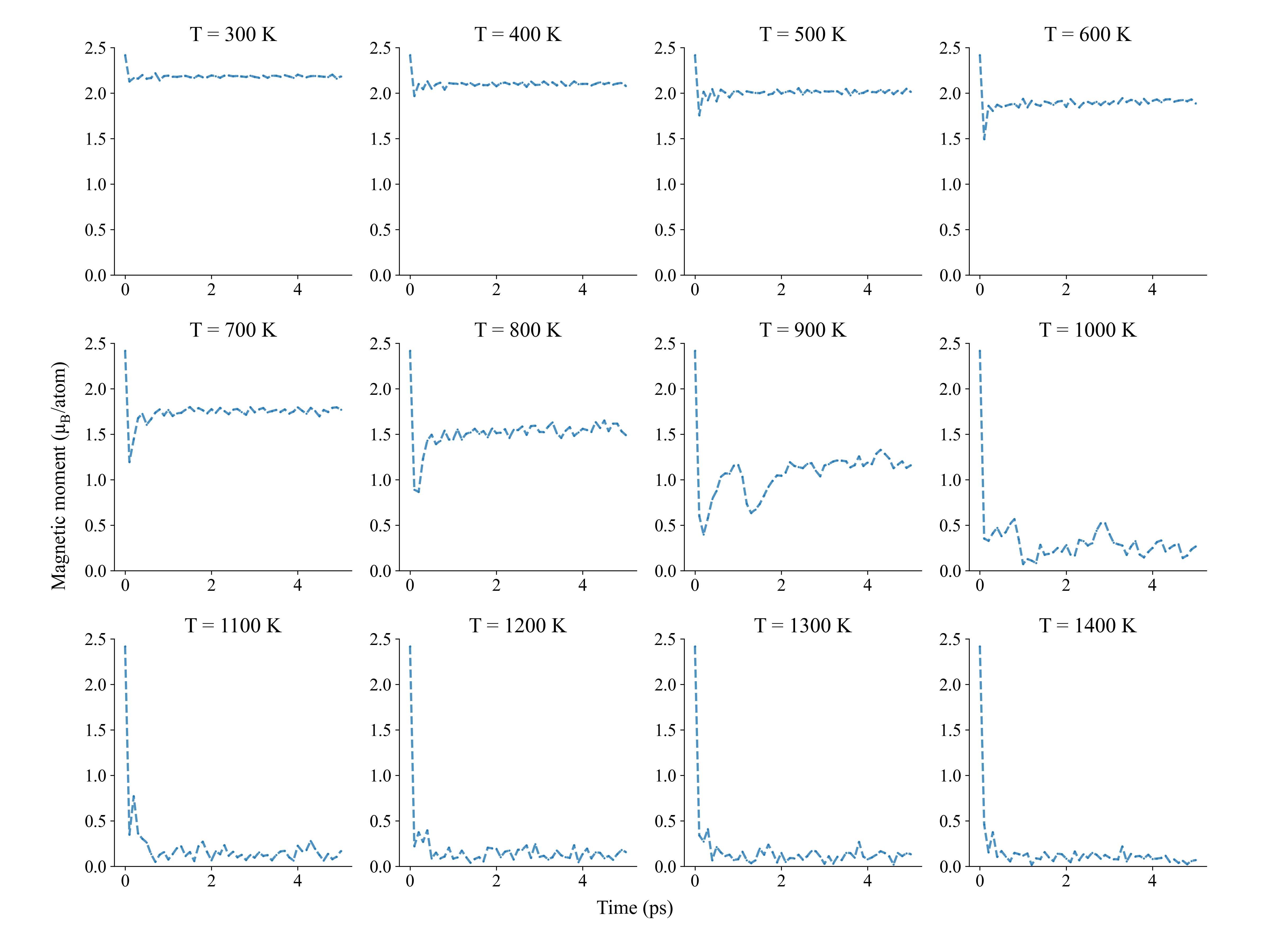}
    \caption{
    \textbf{The evolution of magnetic moments in NVT dynamical simulations over time at different temperatures.}
    We conducted NVT ensemble simulations at different temperatures using the trained models, employing an $8 \times 8 \times 8$ supercell (consisting of 1024 atoms), with a virtual magnetic mass set to 0.01, a timestep of 0.1~fs. 
    }
    \label{fig:}
\end{figure*}

\begin{figure*}
    \centering    
    \includegraphics[width=1.0\textwidth]{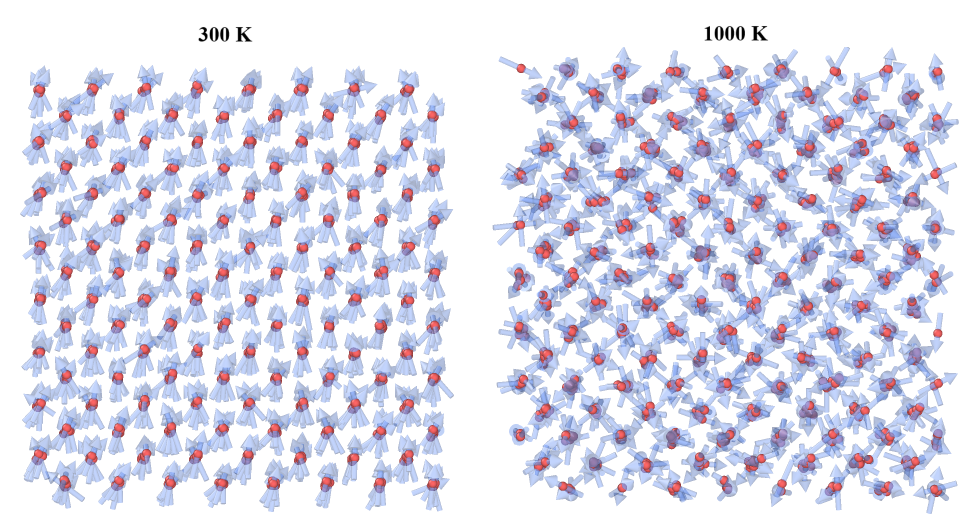}
    \caption{
    \textbf{Ferromagnetic and paramagnetic state.}
    Equilibrated configurations from NVT dynamics simulations at 300 K and 1000 K. Red spheres represent Fe atoms, and blue arrows indicate the magnitude and orientation of atomic magnetic moments.}
    \label{fig:}
\end{figure*}

\begin{figure*}
    \centering    
    \includegraphics[width=1.0\textwidth]{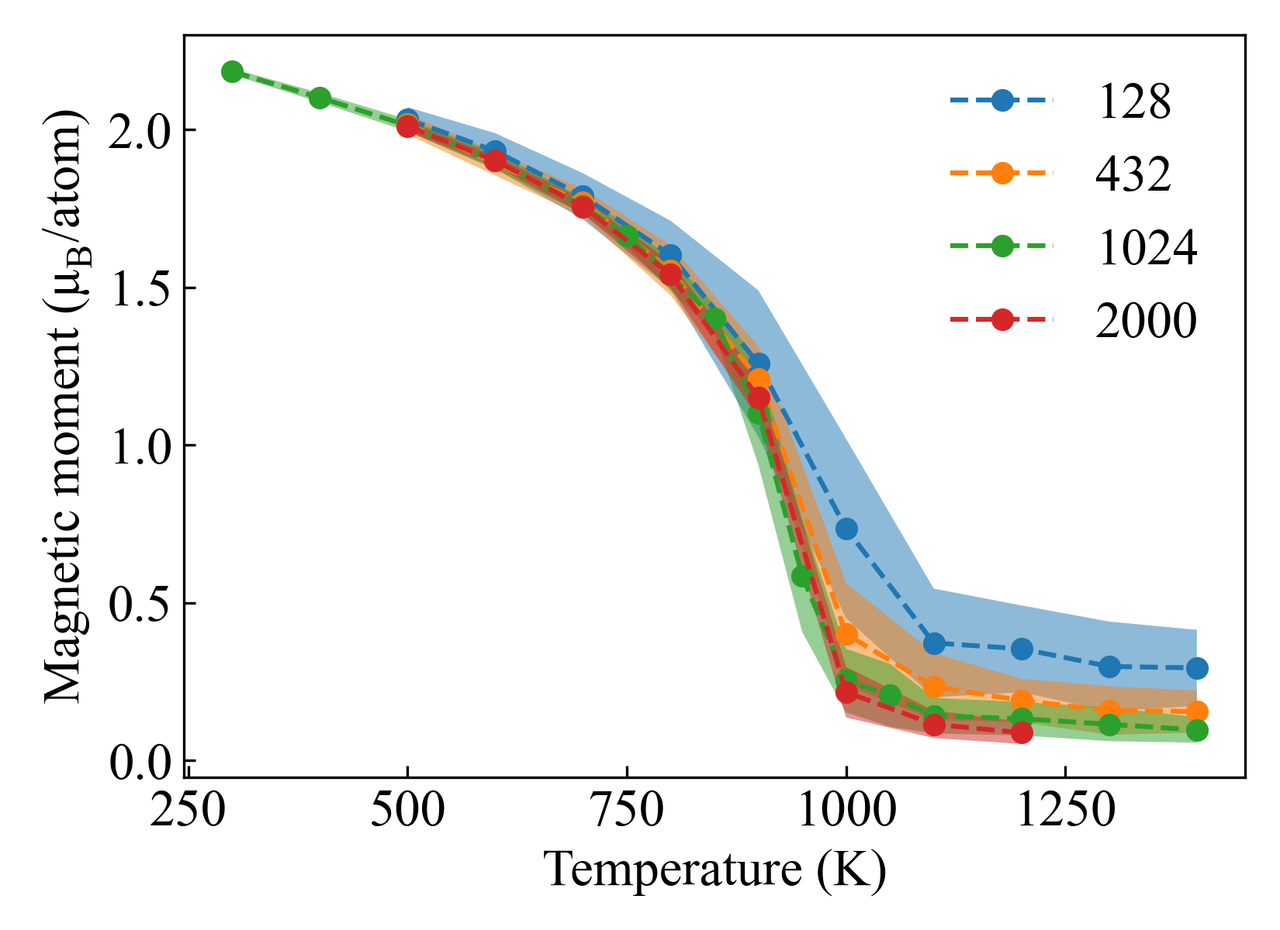}
    \caption{
    \textbf{The size convergence of the model for MD simulations.}
    The total magnetic moment versus temperature obtained from NVT simulations of systems with varying sizes (128/432/1024/2000 atoms). The error bars represent the standard deviation of the total magnetic moment calculated from trajectories at corresponding temperatures.}
    \label{fig:}
\end{figure*}

\begin{figure*}
    \centering    
    \includegraphics[width=1.0\textwidth]{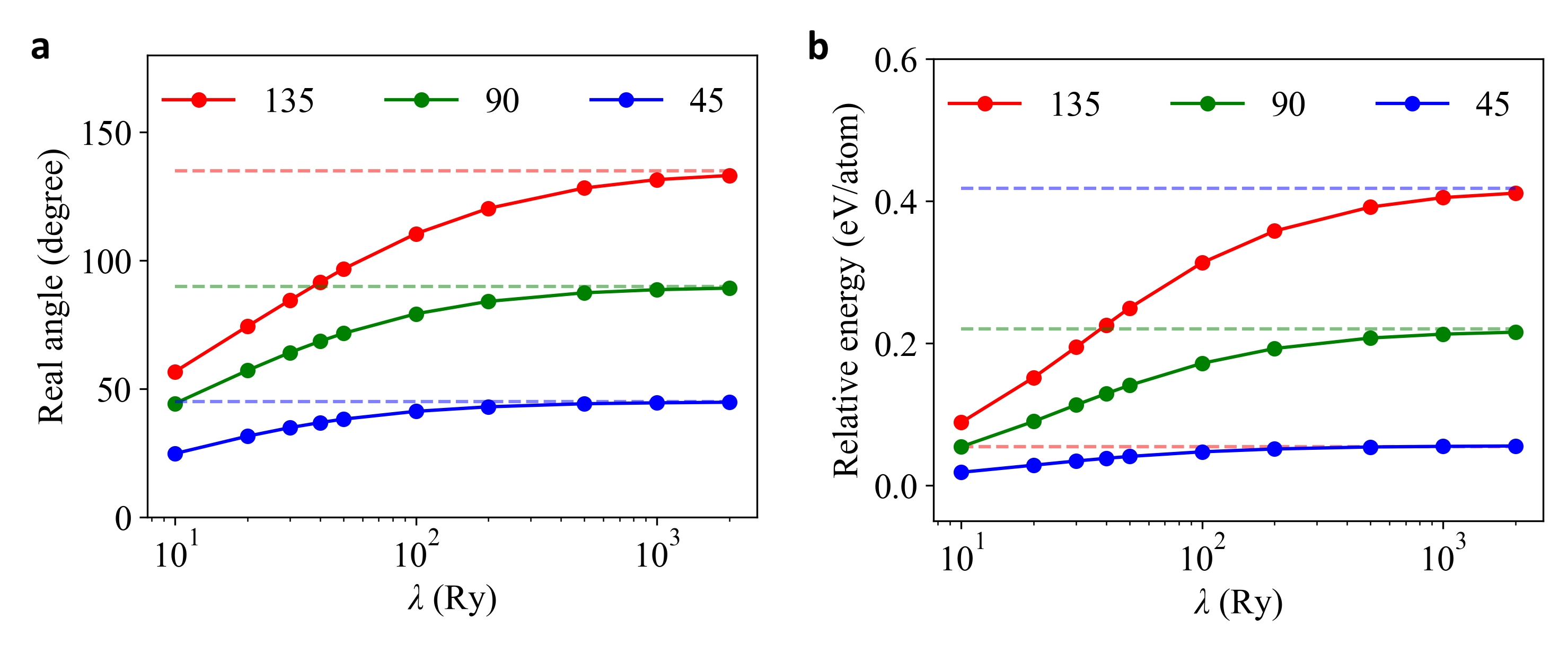}
    \caption{
    \textbf{The comparison between the spin-constrained method in ABACUS and Quantum Espresso 7.0.}
    For the BCC-Fe system, we used ABACUS's cDFT and QE's penalty-function-based cDFT to constrain the magnetic moment angle (45$^\circ$/90$^\circ$/135$^\circ$) between two iron atoms in the unit cell, respectively. (a) The calculated magnetic moment angles in QE under different lambda, with the dashed line indicating the target angle. (b) The total energies under different lambda values (relative to the FM state energy calculated by QE), with the dashed line representing the energy calculated by ABACUS using deltaspin for the target angle (relative to the FM state energy calculated by ABACUS).}
    \label{fig:}
\end{figure*}

\begin{table}[h]
\centering
\caption{
\textbf{Calculation of the z-component DMI strength $D^z_{12}$ between Fe atoms in an Fe-Pt-Fe short chain using the four-states method.} The cDFT approach proposed in this paper serves as a powerful tool for investigating magnetic phenomena related to SOC, as demonstrated through this short example. We consider an Fe-Pt-Fe short chain with an angle of 120 degrees between the two Fe-Pt bonds. Since the system lacks inversion symmetry and Pt, as a heavy element, introduces significant SOC effects that modulate the magnetic interaction between the Fe atoms, it serves as a simple yet pronounced example to showcase SOC effects. First, we performed full relaxation and calculated the ground state of the system. Without introducing cDFT, the magnetic moments on the two Fe atoms converged to a parallel ferromagnetic state, which can be explained by the Heisenberg exchange interaction $J_{ij}\bm{S}_i\bm{S}_j$. However, the SCF calculation fails to capture the Dzyaloshinskii–Moriya interaction (DMI) induced by SOC, $\bm{D}_{ij} \cdot (\bm{S}_i \times \bm{S}_j )$. Here, we employed the conventional four-states method to extract the z-component DMI strength between the Fe atom pair. In the four distinct magnetic configurations, the magnitudes of the magnetic moments on the two Fe atoms remain unchanged, but the spin polarization is fixed along different target directions, while the magnetic moment of the Pt atom remains unchanged. We use cDFT to accurately constrain the magnetic moments to the specified directions and obtain the corresponding total energies $E_1,E_2,E_3,E_4$. The supercell box size was set to at least 15 $\rm{\AA}$ to avoid periodic interaction effects. 
All calculations were performed with the fully-relativistic Dojo pseudopotential and the TZDP basis, incorporating SOC effect. Each inner loop of $\bm{\lambda}$ optimization converges when the maximum atomic magnetism variation is less than $\delta\mathbf{M}<10^{-7} \mu_B$, while the SCF convergence until density difference $\delta\rho<10^{-6}$.
From the energy differences among the four states, we can extract the z-component DMI strength $D^z_{12}$ between the Fe atoms. In this calculation, contributions from other magnetic interactions—such as the Heisenberg exchange coupling, magnetocrystalline anisotropy, and dipole–dipole interaction—are strictly canceled out.}
\label{tab:magnetic_energy}
\begin{tabular}{p{1cm}p{5cm}p{5cm}}
\toprule
Index & Magnetic Configuration & Total Energy (eV) \\
\midrule
1 & $ \bm{S}_{1}=(1,0,0), \bm{S}_{2}=(0,1,0) $ & $E_1$ = -10366.926125 \\
2 & $ \bm{S}_{1}=(1,0,0), \bm{S}_{2}=(0,-1,0) $ & $E_2$ = -10367.004280 \\
3 & $ \bm{S}_{1}=(-1,0,0), \bm{S}_{2}=(0,1,0) $ & $E_3$ = -10367.004277 \\
4 & $ \bm{S}_{1}=(-1,0,0), \bm{S}_{2}=(0,-1,0) $ & $E_4$ = -10366.926134 \\
\midrule
 & & $D^Z_{12} = \frac{(E_1+E_4)-(E_2+E_3)}{4\bm{S}^2} \approx 3.4~\rm{meV}$   \\
\bottomrule
\end{tabular}
\end{table}

\begin{table}[htbp]
  \centering
  \caption{\textbf{Comparison between the total magnetic moment direction and the constrained magnetic moment direction.} Using a BCC-Fe unit cell containing two Fe atoms, we applied constraints at different interatomic angles. The second column displays the direction corresponding to the total magnetic moment obtained by integrating the magnetic density over all real-space grid points. The third column shows the magnetic moment direction determined by the projection method. The fourth column presents the relative error between the two.}
  \label{tab:magnetic_moment_comparison}
  \begin{tabular}{cccc}
    \toprule
    Target interatomic angles (°) & Total magnetism angle (°) & Projection magnetism angle (°) & Error (°) \\
    \midrule
    15  & 7.5018  & 7.5000  & 0.0018  \\
    30  & 15.0491 & 15.0000 & 0.0491  \\
    45  & 22.4751 & 22.4992 & -0.0241 \\
    60  & 30.0488 & 30.0000 & 0.0488  \\
    75  & 37.5590 & 37.5000 & 0.0590  \\
    90  & 45.0000 & 44.9983 & 0.0016  \\
    105 & 52.4808 & 52.4999 & -0.0191 \\
    120 & 60.0101 & 59.9999 & 0.0101  \\
    135 & 67.4916 & 67.5083 & -0.0166 \\
    150 & 75.0228 & 74.9999 & 0.0228  \\
    \bottomrule
  \end{tabular}
\end{table}

\begin{figure*}
    \centering    
    \includegraphics[width=1.0\textwidth]{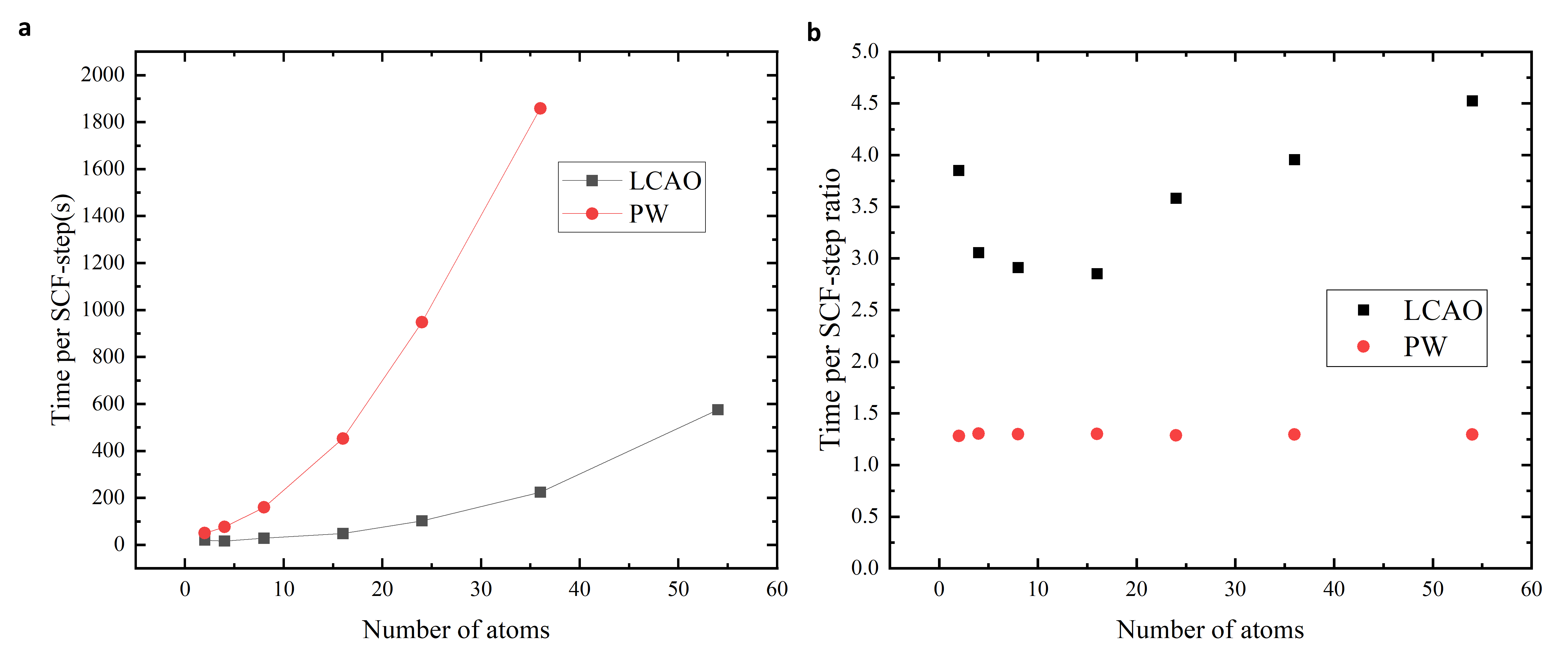}
    \caption{
    \textbf{Tests on the efficiency and scalability of cDFT implementation with PW and LCAO basis sets in ABACUS.}
    Based on the plane-wave (PW) and linear combination of atomic orbitals (LCAO) basis sets available in ABACUS, we constructed supercells of the Fe-BCC system containing different numbers of atoms (2, 4, 8, 16, 24, 36, and 54). For each system, we performed both constrained magnetic calculations—where the magnetic moment angle between two Fe atoms was fixed at 90°—and unconstrained calculations. All computations were executed on identical 16-core CPU nodes.
    (a) illustrates the computational time per SCF step in cDFT calculations as a function of the number of atoms. (b) shows the ratio of the time per SCF step in constrained calculations to that in unconstrained calculations for different system sizes. Specifically, for LCAO, this ratio ranges from 3 to 5, reflecting the fact that its inner-loop solver employs the same diagonalization method as the outer loop. Thus, the ratio effectively represents the average number of iterations required in the inner loop. In contrast, the computational overhead associated with the PW basis set is significantly smaller, adding only about 30\% to the total computation time. This efficiency stems from the use of a more effective algorithm for estimating magnetic moments in the inner loop, as noted in the main text. Furthermore, for both PW and LCAO, the additional computational cost due to constraints remains nearly constant as the system size increases. This indicates that the constraints do not alter the scaling behavior of the SCF calculations, but rather increase the prefactor in the computational cost.
    }
    \label{fig:}
\end{figure*}

\begin{table}[h]
\centering
\caption{
\textbf{Comparison of convergence and efficiency between ABACUS-PW's cDFT method and QE v7.0's Penalty-based method.}
We randomly selected 10 cDFT calculations from the tests in Fig. S14 that were performed using QE for analysis. For comparison, we conducted cDFT calculations under the same constraint angle settings using ABACUS. Columns 2-4 present the number of converged SCF steps (SCF Steps), total wall time (Total Time), and time per step (Time Per Step) for ABACUS and QE, respectively, with the values for QE indicated in parentheses. The fifth column shows the ratio of the time per step between the two methods. To minimize the influence of other factors, we ensured that all parameters unrelated to the cDFT settings were consistent between the two software packages, including the computational node, parallelization settings, pseudopotentials, energy cutoff, mixing parameters, convergence criteria, etc. All calculations were performed using a 32-core CPU with MPI parallelization, and the Davidson method for diagonalization.
}
\begin{tabular}{cccccc}
\toprule
Constraint angle (°) & SCF Steps & Total Times (s) & Time Per Step (s) & \begin{tabular}{@{}c@{}}ABACUS Time Per Step/ \\ QE Time Per Step\end{tabular} \\
\midrule
33 & 8 (26) & 195.14 (529.29) & 24.39 (20.36) & 1.20 \\
37 & 8 (33) & 202.43 (663.87) & 25.30 (20.12) & 1.26 \\
42 & 9 (33) & 209.34 (622.24) & 23.26 (18.86) & 1.23 \\
54 & 9 (32) & 209.82 (617.89) & 23.31 (19.31) & 1.21 \\
69 & 9 (27) & 211.43 (549.98) & 23.49 (20.37) & 1.15 \\
74 & 8 (26) & 190.70 (510.68) & 23.84 (19.64) & 1.21 \\
80 & 9 (39) & 208.00 (743.19) & 23.11 (19.06) & 1.21 \\
88 & 9 (32) & 215.37 (602.29) & 23.93 (18.82) & 1.27 \\
117 & 9 (25) & 220.52 (500.74) & 24.50 (20.03) & 1.22 \\
125 & 10 (24) & 245.53 (465.65) & 24.55 (19.40) & 1.27 \\
\midrule
average & 8.8 (29.7) & 210.8 (580.5) & 23.9 (19.6) & 1.22 \\
\bottomrule
\end{tabular}
\end{table}